\documentclass[12pt,preprint]{aastex}
\begin{document}
\setcounter{figure}{0}
\title{Proper Motions of the LMC and SMC: Reanalysis of \textit{Hubble
Space Telescope} Data\footnote{Based on observations with NASA/ESA
\textit{Hubble Space Telescope}, obtained at the Space Telescope
Science Institute, which is operated by the Association of
Universities for Research in Astronomy, Inc., under NASA contract NAS
5-26555.}}

\author{Slawomir Piatek} \affil{Dept. of Physics, New Jersey Institute
of Technology,
Newark, NJ 07102 \\ E-mail address: piatek@physics.rutgers.edu}

\author{Carlton Pryor}
\affil{Dept. of Physics and Astronomy, Rutgers, the State University
of New Jersey, 136~Frelinghuysen Rd., Piscataway, NJ 08854--8019 \\
E-mail address: pryor@physics.rutgers.edu}

\author{Edward W.\ Olszewski}
\affil{Steward Observatory, University of Arizona,
    Tucson, AZ 85721 \\ E-mail address: eolszewski@as.arizona.edu}

\begin{abstract}

Kallivayalil et al.\ have used the \textit{Hubble Space Telescope} to
measure proper motions of the LMC and SMC using images in 21 and five
fields, respectively, all centered on known QSOs.  These results are
more precise than previous measurements, but have surprising and
important physical implications: for example, the LMC and SMC may be
approaching the Milky Way for the first time; they might not have been
in a binary system; and the origin of the Magellanic Stream needs to
be re-examined.  Motivated by these implications, we have reanalyzed
the original data in order to check the validity of these
measurements.  Our work has produced a proper motion for the LMC that
is in excellent agreement with that of Kallivayalil et al., and for
the SMC that is in acceptable agreement.

We have detected a dependence between the brightness of stars and their
mean measured motion in a majority of the fields in both our reduction
and that of Kallivayalil et al.  Correcting for this systematic error
and for the errors caused by the decreasing charge transfer efficiency
of the detector produces better agreement between the measurements from
different fields.  With our improved reduction, we do not need to
exclude any fields from the final averages and, for the first time
using proper motions, we are able to detect the rotation of the LMC.
The best-fit amplitude of the rotation curve at a radius of 275~arcmin
in the disk plane is $120 \pm 15$~km~s$^{-1}$.  This value is larger
than the 60--70~km~s$^{-1}$ derived from the radial velocities of HI
and carbon stars, but in agreement with the value of 107~km~s$^{-1}$
derived from the radial velocities of red supergiants.

Our measured proper motion for the center of mass of the LMC is
$(\mu_{\alpha}, \mu_{\delta}) = (195.6\pm 3.6, 43.5 \pm 3.6)$~mas
century$^{-1}$; that for the SMC is $(\mu_{\alpha}, \mu_{\delta}) =
(75.4\pm 6.1, -125.2 \pm 5.8)$~mas century$^{-1}$.  The uncertainties
for the latter proper motion are 3 times smaller than those of
Kallivayalil et al.

\end{abstract}


\section{Introduction}
\label{intro}

The Magellanic Clouds span many degrees on the sky owing to their
relatively large size and proximity to the Milky Way (heliocentric
distances are 50~kpc for the LMC and 62~kpc for the SMC).  The LMC is
the most luminous among the satellite dwarf galaxies of the Milky Way.
With nascent spiral arms and a bar, the LMC is a late-type spiral rich
in gas and with active star formation.  Spectroscopic studies of the
galaxy show a sizeable rotation \citep[\textit{e.g.},][]{om07}.  In
contrast, the SMC is a dwarf irregular with less active star formation
and a smaller and still poorly-measured rotation.  The LMC and SMC are
close together on the sky and are connected in projection by a bridge
of HI.  The Magellanic Stream, an approximately $100^\circ$-long
distribution of HI, extends from the HI around the Clouds.  A second
stream containing less HI emanates in the opposite direction
\citep[][]{p98,b05}.  The apparent gaseous bridge between the LMC and
SMC could have arisen from an interaction between these two galaxies,
and modeling has suggested that they may have been or may be a bound
pair \citep[\textit{e.g.},] []{gn96}.  There is a long-standing
interest in understanding the relations between the LMC, SMC, and
Stream \citep[\textit{e.g.},][]{m74, p03, b07, ni07}.

Because of the proximity of the two galaxies to the Milky Way, it is
intriguing to speculate that the tidal field of the Milky Way has had
a significant impact on the evolution of the Magellanic Clouds.  For
example, the Stream is widely considered to consist of gas removed
from the LMC or SMC by a combination of ram pressure and tidal
interaction with the Milky Way.  Such an origin of the Stream implies
that it shares the orbital plane of the LMC or SMC.  Among the several
quantities needed to answer the above questions, the proper motions
are crucial since, together with the radial velocities and distances,
they give the current space velocities of the galaxies.  These
velocities are necessary initial conditions in determining the past or
the future orbits for a given Galactic potential.  Alternatively,
modeling the Magellanic Stream may constrain the potential of the
Galaxy if the proper motions of the LMC and SMC are known with
sufficient precision \citep[\textit{e.g.},][] {hr94,l95}.

Recognizing the importance of the proper motions of the Magellanic
Clouds, several groups have attempted to measure them.  In
chronological order, the measurements for the LMC are: \citet{j94},
\citet{k94}, \citet{kb97}, \citet{d01}, \citet{pe02}, \citet{pe06},
and \citet[][K06a]{k06a}.  For the SMC, the measurements are: \citet{kb97},
\citet{i99}, and \citet[][K06b]{k06b}.

The measurements by K06a and K06b used images taken with the
\textit{Hubble Space Telescope} (HST) and they have uncertainties that
are only one-third as large as those of the best previous
measurements.  Each of the 21 fields in the LMC and five in the SMC
has a confirmed QSO which serves as a standard of rest.  The analysis
is based on the methodology developed by \citet{ak04}.  Similar data
and analyses have measured proper motions for dwarf spheroidal
companions of the Milky Way with comparable uncertainties to those in
K06a and K06b
\citep[\textit{e.g.},][]{p07}.  The proper motion for the LMC reported
by K06a is $(\mu_{\alpha}, \mu_{\delta})= (203 \pm 8, 44 \pm
5)$~mas~century$^{-1}$ and by K06b for the SMC is $(\mu_{\alpha},
\mu_{\delta})= (116 \pm 18, -117 \pm 18)$~mas~century$^{-1}$.  These
values yield large space motions which then imply that, for example,
the LMC and SMC may be on their first approach to the Milky Way, that
the LMC and SMC may not initially have been bound to each other, and
that models for the formation of the Magellanic Stream via an
interaction with the Milky Way need to be re-examined \citep[see]
[]{b07, ni07}.  Thus, an independent check of the results in K06a and
K06b is worth having and this article reports on a reanalysis of their
data.  Section~\ref{sec:data} describes the data;
section~\ref{sec:mpm} explains the process of deriving the proper
motion using our method; section~\ref{sec:res} presents our results
and compares them to those in K06a and K06b; section~\ref{sec:disc}
discusses the implications of the measured proper motions; and
section~\ref{sec:summary} is a summary of the main results.

\section{Observations and Data}
\label{sec:data}

The data consist of images in the F606W and F814W bands obtained with
the High Resolution Camera (HRC) of the Advanced Camera for Surveys
(ACS).  The images were produced by the ACS data-reduction pipeline
and provided by the Space Telescope Science Institute archive; these
are the same data as those used by K06a and K06b.  The images were
taken in a snapshot mode at two epochs.  In the case of the LMC, 21
fields are common to both epochs and, in the case of the SMC, five are
common to both.  The time between epochs ranges from 1.1 to 2.8 years.
Each field is centered on a confirmed QSO.  Almost all of the pairs of
images have orientations (i.e., the HST ORIENTAT angle) differing by
tens of degrees between the epochs.  For comprehensive information
about the observations and data, see Table~1 and Figure~1 in K06a for
the LMC and in K06b for the SMC.

\section{Measuring Proper Motion}
\label{sec:mpm}

A series of articles beginning with \citet{p02} describe our basic
technique for deriving proper motions.  Central to our method is the
presence of a QSO in each observed field which serves as an
extragalactic ``reference point.''  The crucial steps of the method
are: 1. Derive an effective point-spread function
\citep[ePSF;][]{ak00} at each epoch using stars and the QSO in
dithered images.  Our experience shows that the PSF for a QSO is
similar to that for a star, making the bright, compact QSO an ideal
reference point.  2. Determine accurate centroids for the stars and
the QSO by fitting the ePSF.  3.  Correct the centroids for the known
geometrical distortions in the camera and CCD.  4. Transform the
centroids of stars and the QSO measured at different epochs to a
common coordinate system which moves together with the stars of the
galaxy.  For the QSO and those stars that are not members of the
galaxy, a fitted linear motion is included in the coordinate
transformation.  The proper motion of the galaxy derives from the
motion of the QSO.

When deriving the transformation to a common coordinate system, a
linear motion is always fitted for the QSO.  A motion is also fitted
for objects whose contribution to the total $\chi^2$ of the scatter
around the transformation is above 9.21, the value which should be
exceeded by chance only 1\% of the time.  Except for the QSO, the
objects with fitted motion are likely to be foreground stars of the
Milky Way.  Once the parameters of the transformation are determined,
the motion of each remaining object without a fitted motion is
calculated from the transformed coordinates at each epoch; this motion
should be zero within its uncertainty.

Our method uses the most general linear transformation, which contains
six fitted parameters, between the coordinate systems at different
epochs.  Plots of position residuals
\textit{versus} location on the CCD showed that more parameters were
unnecessary.  The transformation also corrects for the effects caused
by the degrading charge transfer efficiency (CTE) of the CCD in the
HRC \citep[see][]{br05}.  The method used is similar to that in
\citet{p05} and \citet{p07}: the $Y$ coordinate of an object is
corrected by an amount that depends on the brightness of the object
and is linearly proportional to $Y$ and to the time since ACS was
installed.  This last dependence is supported by the evidence provided
in the ACS Handbook \citep{pav06}.  We adopted a correction that
varies with the $S/N$ of the object as $(S/N)^{-0.42}$ between a $S/N$
of 10 and 100 and is constant at the boundary values outside of that
range.  The exponent also comes from data in the ACS Handbook.  The
final proper motions do not depend sensitively on the details of how
the CTE corrections are made.  The above method depends on a single
parameter, which is the rate of change with time of the correction
applied to the $Y$ coordinate of an object with a $S/N$ of 15 at a $Y$
location of 1024~pixels.  Fitting for this parameter using some of the
data least affected by the systematic errors discussed below indicated
a value of $0.030$~pixel~yr$^{-1}$.  All results reported in this
article used corrections calculated with this value.

K06a and K06b did not make corrections as a function of stellar flux
for the shifts in centroids due to degrading CTE.  With many
independent fields in the LMC, K06a argue that the effect of these shifts
on the average proper motion approaches zero as
$N^{-1/2}$, provided that the $N$ fields have
an isotropic distribution of position angles.  However, the effect on
the proper motion of the SMC may be greater because there are only five
fields and the distribution of image orientations is not isotropic (four
of these fields have similar HST ORIENTAT angles at the first epoch).

\clearpage

\begin{figure}[h]
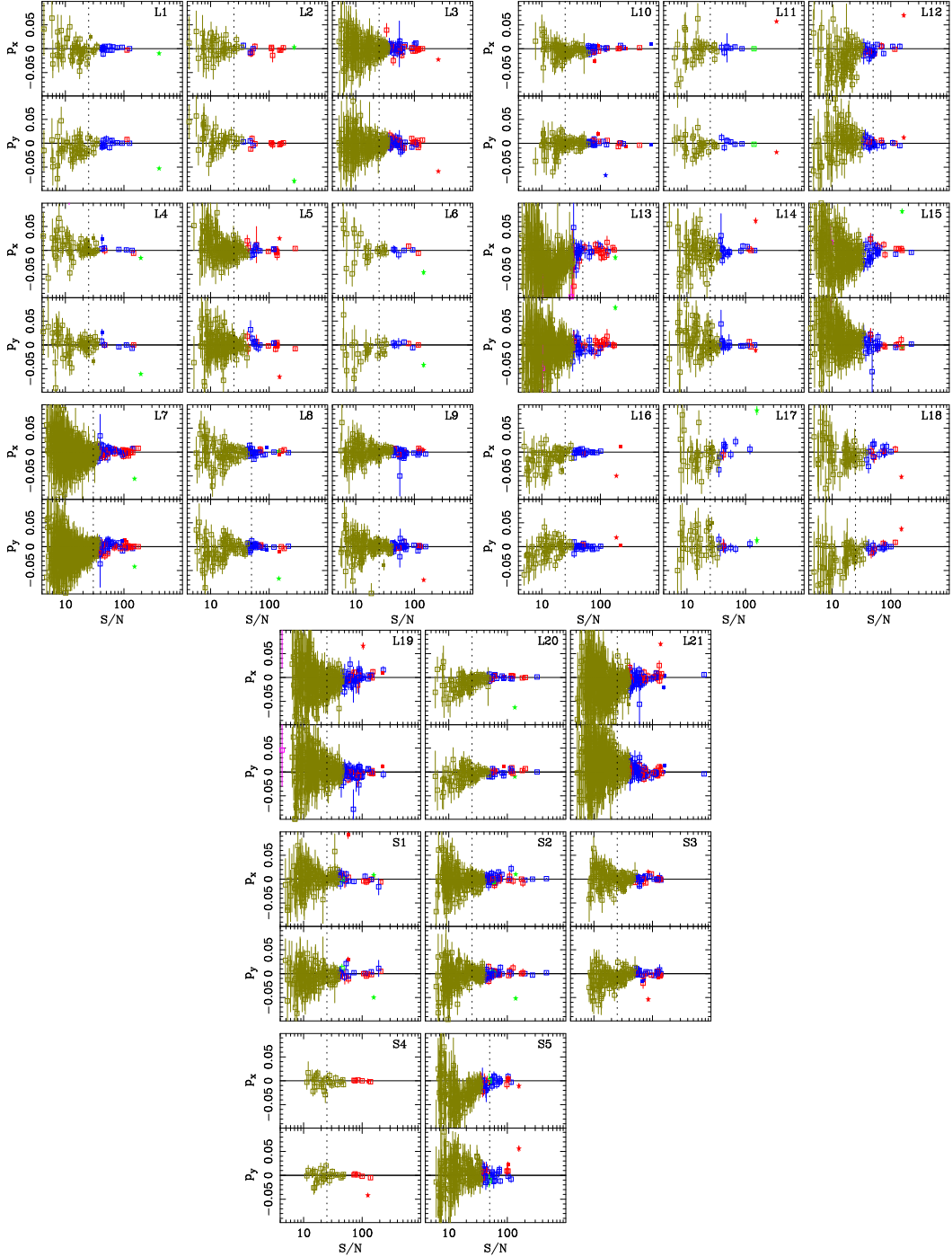

\centering
\includegraphics[angle=0,scale=0.35,clip=true]{f1a.eps}
\includegraphics[angle=0,scale=0.35,clip=true]{f1b.eps}
\includegraphics[angle=0,scale=0.35,clip=true]{f1c.eps}
\caption{Motion in the
common coordinate system, $p_{x}$ and $p_{y}$ in pix~yr$^{-1}$,
\textit{versus} $S/N$ for the 21 fields in the LMC and five in the
SMC.  The points are color-coded depending on the location of objects
in their respective CMDs, which are shown in Figure~\ref{fig:cmd}.
Star symbols represent the QSOs and squares represent the stars.
Filled squares correspond to those stars that have fitted motion.
Note the trends between $p_{x}$ or $p_{y}$ with $S/N$ for a majority
of fields, e.g. L13.  To reduce the impact of these trends on the
proper motion, only objects with $S/N$ greater than the value
indicated by a vertical dashed line were used in fitting for a
transformation.  Column 2 of Tables~1 and 3 gives these values of
$S/N$ for the fields in the LMC and SMC, respectively.}
\label{fig:pmsn}
\end{figure}

\begin{figure}[t]
\centering
\includegraphics[angle=0,scale=0.75,clip=true]{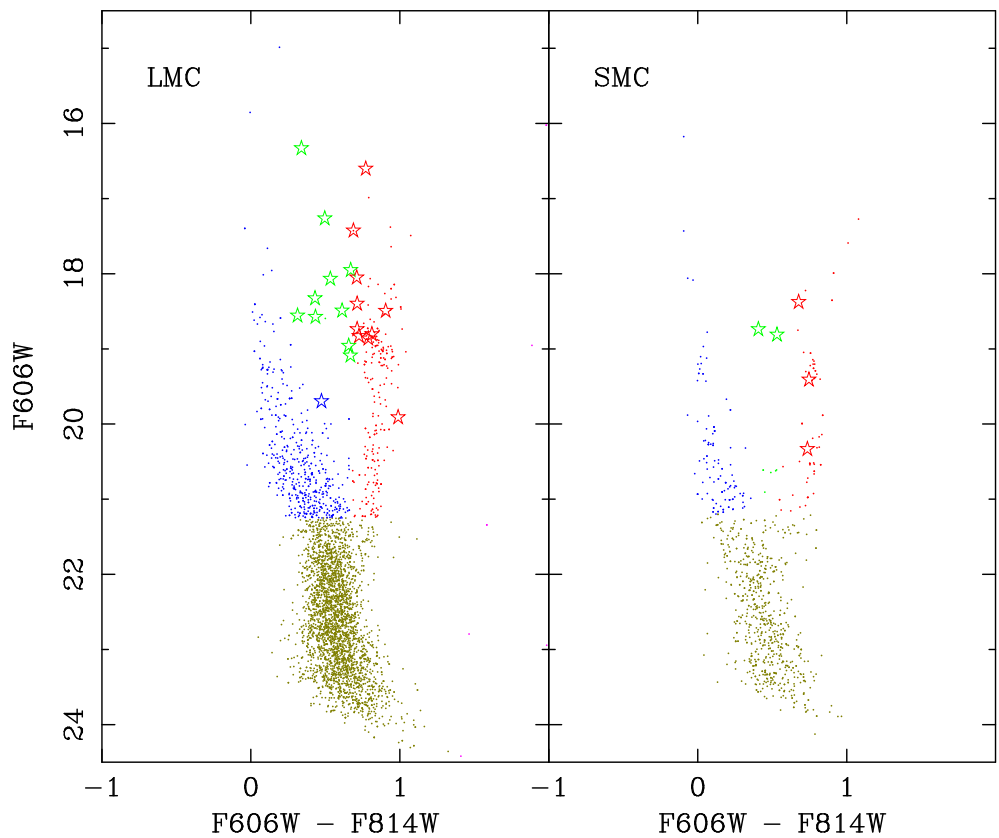}
\caption{Color-magnitude diagrams for the LMC \textit{left panel} and
the SMC \textit{right panel}.  The diagrams show only those objects
that were matched at the two epochs and, thus, whose motion in the
common coordinate system can be determined.  The QSOs are marked with
a star symbol.  The points are color-coded depending on their location
in the diagram.  No corrections for reddening or extinction were
applied.}
\label{fig:cmd}
\end{figure}

\clearpage

To examine the effect of degrading CTE on our data, we plotted the
motions in the common coordinate system, $p_{x}$ and $p_{y}$ in
pix~yr$^{-1}$, of all objects \textit{versus} their $S/N$.
Figure~\ref{fig:pmsn} shows these plots for all of the fields in the
LMC and SMC.  A majority of the fields show trends in these plots,
particularly for $S/N$ less than about 20.  However, these trends were
sometimes along the direction orthogonal to that expected from a degrading
CTE and sometimes in the expected direction, but with the opposite of
the expected sign.  None of the large trends were well removed by fitting
our model for the CTE correction.  We
conclude that these trends arise from some effect other than the
degrading CTE.  A possible explanation could be an error in the ePSF,
but varying the parameters used in the construction of the ePSF had no
effect on the trends.  As we discuss below, these trends are likely to
be present in the results of K06a and K06b too.  To minimize the
effect of the dependence of mean motion on $S/N$, we limit the sample
of stars used to determine the transformation between epochs to stars
with $S/N$ above a limit that is usually 25 but can be as large as 50.
These limits are indicated in Figure~\ref{fig:pmsn} by vertical dashed
lines and they are also listed in column (2) of Tables 1 (for the LMC)
and 3 (for the SMC).  The limit is chosen empirically so that the mean
motion of stars with a $S/N$ similar to that of the QSO is zero.  A
concern is that a change in the PSF with color, which has not been
modeled in either analysis, is causing the observed trends.  The
points in the plots depicted in Figure~\ref{fig:pmsn} are color-coded
depending on the location of objects in their respective
color-magnitude diagrams (CMDs), which are shown in
Figure~\ref{fig:cmd}.  The photometry for each CMD was derived using
HSTPhot \citep{do00} from the first-epoch images taken in the F606W
and F814W filters and has not been corrected for reddening and
extinction.  Visual inspection of Figure~\ref{fig:pmsn} does not
provide evidence for a systematic difference between the mean motions
of red and blue stars at high $S/N$.  While fields such as L13 hint at
such a difference, the majority of the fields do not.  To
quantitatively estimate the size of any possible color effect, we
calculated separately the weighted mean motion for the red and blue
stars with $S/N > 100$ and located in all of the fields in the LMC.
The resulting differences in the $X$ and $Y$ directions between the
weighted mean motions for the red and blue stars are $-1.2\times
10^{-4} \pm 6.6\times 10^{-4}$~pix~yr$^{-1}$ and $3.4\times 10^{-4}
\pm 6.2\times 10^{-4}$~pix~yr$^{-1}$, respectively.  Both differences
are consistent with zero within their uncertainties and, thus, the
measured motion of the QSO in our derived common coordinate system is
an accurate reflection of the motion of the LMC or the SMC.

\clearpage

\begin{figure}[h]
\centering
\includegraphics[angle=-90,scale=0.75]{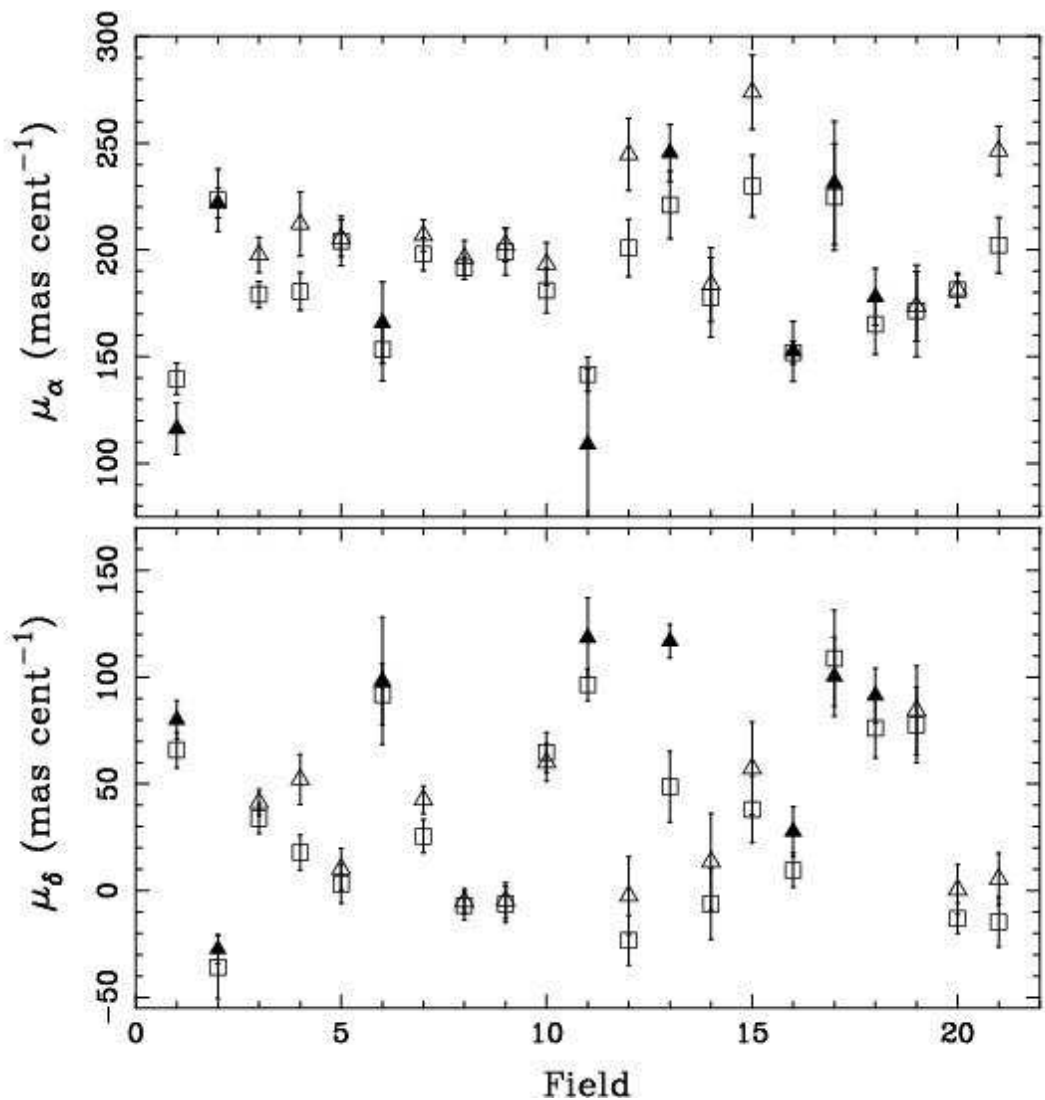}
\caption{Comparison of measured proper motions for the LMC.  Squares
represent the values reported by this article, whereas triangles
represent those in K06a.  Both sets of values are from Table~1.  Solid
triangles correspond to those fields that were excluded in
the calculation of the mean proper motion in K06a.  \textit{Top
panel}:  $\mu_{\alpha}$ \textit{versus} field number.  \textit{Bottom
panel}: $\mu_{\delta}$ \textit{versus} field number.  Both panels have
the same vertical scale.}
\label{fig:LMCpmr}
\end{figure}

\clearpage

\section{Results}
\label{sec:res}

\begin{figure}[h!]
\centering
\includegraphics[angle=-90,scale=0.75]{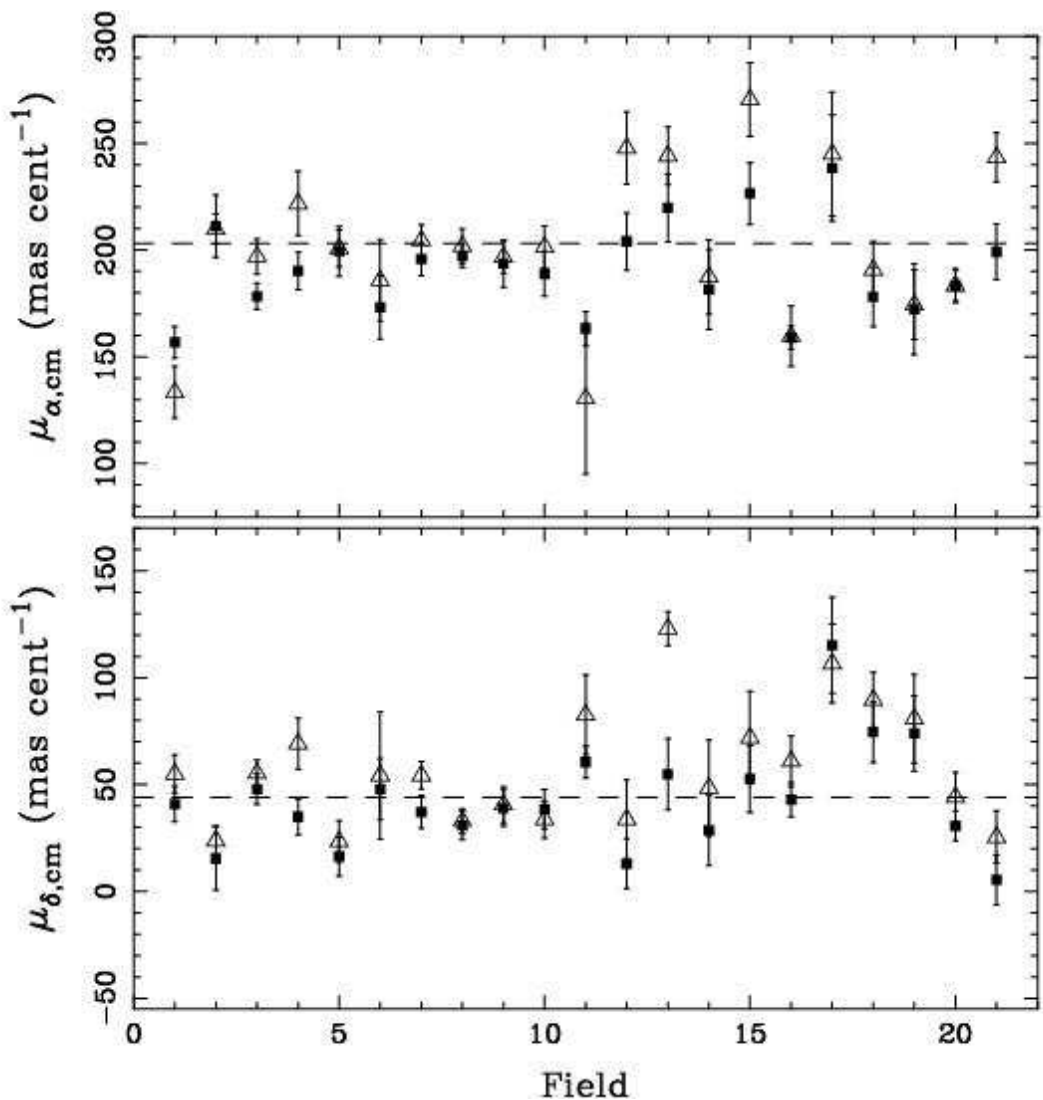}
\caption{Comparison of center-of-mass proper motions for the LMC.
Filled squares represent the values reported by this article, whereas
triangles represent those in K06a.  Both sets of values are from
Table~2.  The corrections for rotation and changing perspective are
from K06a.  \textit{Top panel}: $\mu_{\alpha}$ \textit{versus} field
number.  \textit{Bottom panel}: $\mu_{\delta}$ \textit{versus} field
number.  The dashed horizontal lines are mean proper motions for each
component from K06a.  Both panels have the same vertical scale, which
is also the same as in Figure~\ref{fig:LMCpmr}.}
\label{fig:LMCpmc}
\end{figure}

\begin{figure}[t]
\centering
\includegraphics[angle=-90,scale=0.75]{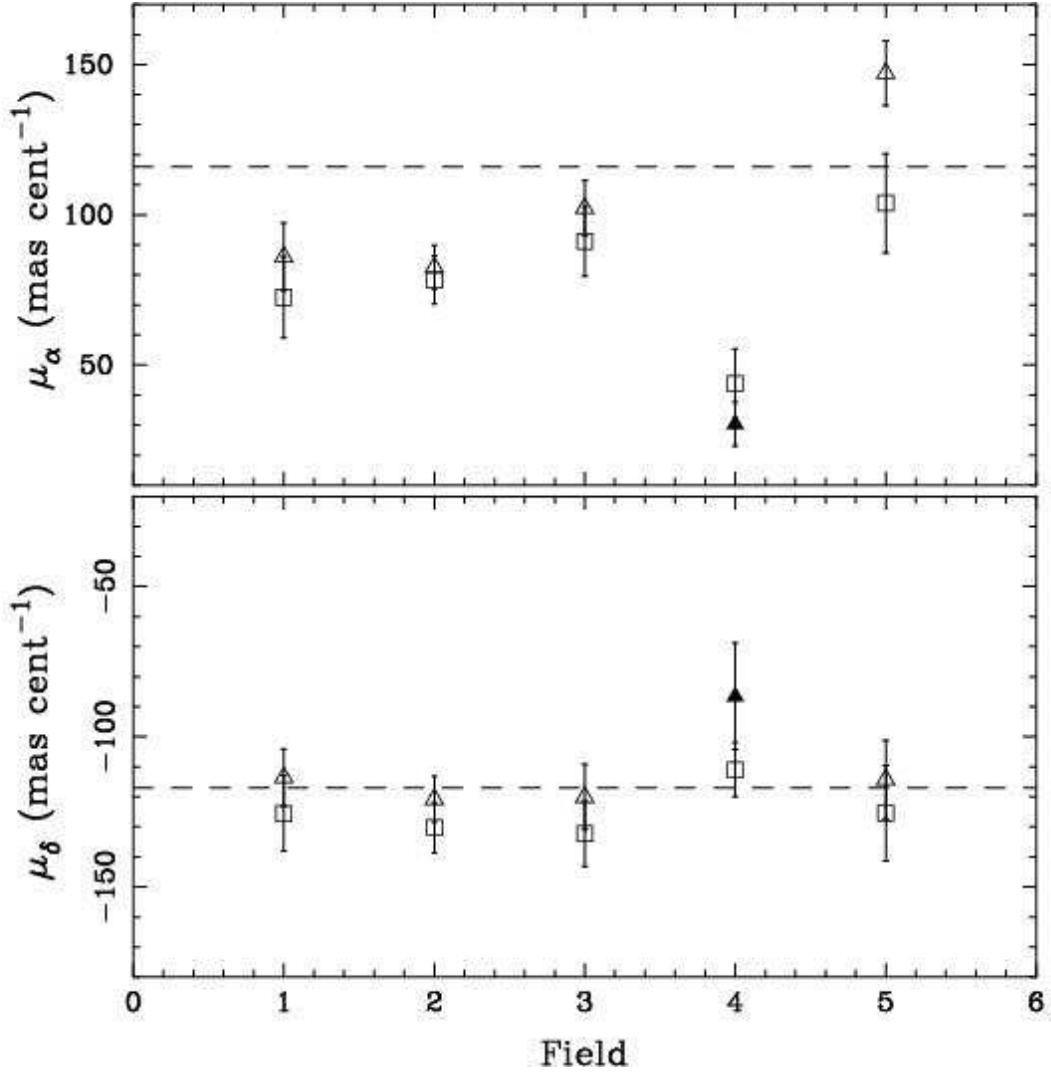}
\caption{Comparison of measured proper motions for the SMC.  Squares
represent the values reported by this article, whereas triangles
represent those in K06b.  Both sets of values are from Table~3.  The
solid triangle corresponds to the field that was excluded in the
calculation of the mean proper motion in K06b.  The dashed horizontal
lines are mean proper motions for each component from K06b.
\textit{Top panel}: $\mu_{\alpha}$ \textit{versus} field number.
\textit{Bottom panel}: $\mu_{\delta}$ \textit{versus} field number.
Both panels have the same vertical scale.}
\label{fig:SMCpmr}
\end{figure}

We have derived proper motions for all 21 fields in the LMC (L1 ---
L21) and all five fields in the SMC (S1 --- S5).  Table~1 provides a
side-by-side comparison of our results for the LMC with those of K06a.
Column~(1) gives the name of a field, column~(2) gives the $S/N$
limit, and column~(3) gives the resulting number of stars used in
fitting the transformation between epochs.  Columns (4) and (5) give
the components of the measured proper motion derived by us in the
equatorial coordinate system, whereas columns (6) and (7) do the same
for the proper motions in K06a.  Columns (8) and (9) are the
difference between our results and those in K06a.  The listed
uncertainty for a difference is the sum in quadrature of the
uncertainties in the two values, even though this uncertainty
indicates the difference expected between two independent measurements
rather than the difference arising from different methods of analyzing
the same data.  Figure~\ref{fig:LMCpmr} plots the components of the
proper motions in columns (4) --- (7) \textit{versus} field number.
The $\mu_{\alpha}$ values are in the top panel and the $\mu_{\delta}$
values are in the bottom.  Squares are our values and triangles are
those in K06a.  Filled triangles are those measurements in K06a that
were not used in their calculation of the average proper motion.  In
the LMC, the difference between the observed proper motion for a field
and the proper motion of the center of mass is significant because of
the changing perspective of the space velocity and the internal
rotation.  Thus, Table~2 lists and Figure~\ref{fig:LMCpmc} plots the
values from Table~1 corrected for these effects.  The corrections are
from K06a.  For the SMC, these corrections are negligible.  Table~3
and Figure~\ref{fig:SMCpmr} compare our results for the SMC with those
of K06b.

\clearpage

\clearpage

The agreement between our results and those in K06a and K06b is good
in most cases.  Because the data in the two studies are the same, any
differences are due to the methods of analysis.  The bottom two lines
of both Table~1 and Table~3 give the mean difference and \textit{rms}
scatter between our results and those of K06a and K06b.  We give the
mean instead of the weighted mean because, as noted above, the listed
uncertainties are not directly related to the size of the differences.
The means of the differences are, for the LMC, comparable to the
uncertainty in the galaxy proper motion given by K06a and, for the
SMC, are smaller.  Tables~1 and 3 show that ten fields in the LMC (L1,
L3, L4, L7, L11, L12, L13, L15, L16, and L21) and two fields in the
SMC (S4 and S5) have differences in at least one component that are
larger than the listed uncertainty.  Most of these twelve fields show
trends of the mean measured motion with $S/N$ and the field with the
largest difference, L13, has one of the largest trends (see
Figure~\ref{fig:pmsn}).  For these twelve fields, reducing the $S/N$
limit from our adopted values, \textit{i.e.},
including more of the stars in the transformation, makes our measured
proper motions closer to the values found by K06a.  Thus, we conclude
that the K06a and K06b results for these twelve fields are affected by
the same systematic errors that depend on $S/N$.  K06a and K06b rejected
from their samples individual stars with discrepant or uncertain proper
motions, but this will not necesarily eliminate a systematic error that
affects all of the stars with the same brightness similarly.  The systematic
error does not appear in their plot of the amplitude of the stellar proper
motions \textit{versus} magnitude because the data from all of the fields
are shown together and the different fields have different trends.
K06a note that their fitted
transformation between epochs for field L13 had an unusually large
$\chi^2$ per degree of freedom, leading them to reject this field from
their average despite it containing one of the largest samples of
stars.  Also rejected were fields with 16 or fewer stars in the final
sample and this tends to elminate fields whose mean proper motion could
be strongly affected by the systematic error.  Thus, the procedures
adopted by K06a and K06b tended to limit the effect of the systematic
error on their final result.  However,
figures~\ref{fig:LMCpmc} and \ref{fig:SMCpmr} show that our
values derived with a $S/N$ limit of 25 or higher make the proper
motions of the twelve fields more consistent with those of the other
fields.

\clearpage

\clearpage

K06a and K06b identified several ``low-quality'' fields, marked with
solid triangles in Figures~\ref{fig:LMCpmr} and \ref{fig:SMCpmr}, on
the basis of small sample size or a large $\chi^2$ per degree of
freedom and excluded them from the calculation of the mean proper
motion.  Most, though not all, of these fields had poor agreement with
the mean proper motion (see Figures~\ref{fig:LMCpmc} and
\ref{fig:SMCpmr}).  After removing the effects of trends with $S/N$, we
find no indication of serious problems at any stage of the analysis for
all 26 fields.  Thus, we conclude that all of the fields contain useful
information about the motions of the LMC and SMC.  Some fields do
deviate from the mean proper motion by more than is expected on the
basis of their uncertainties, most notably L1, L11, L16, L17, and S4.
These fields are likely providing information about internal motions in
the LMC and SMC, and we test for such motions in Sections~\ref{sec:LMC}
and \ref{sec:SMC}.

\section{Discussion}
\label{sec:disc}

Numerous factors can influence the internal motions of a galaxy.  The
distribution of mass with radius determines the shape and amplitude of
the rotation curve in a disk system or the dependence of velocity
dispersion on radius in a pressure-supported system.  The presence of a
bar or a strong tidal disturbance can induce their own streaming
motions.  Old and young stellar populations can have distinct
kinematics, as is well known in the case of the Milky Way.  Below we
discuss what information the measurements of the proper motions in the
LMC and SMC contain about internal motions.

\subsection{The LMC}
\label{sec:LMC}

\begin{figure}[t!]
\centering
\includegraphics[angle=0,scale=0.9]{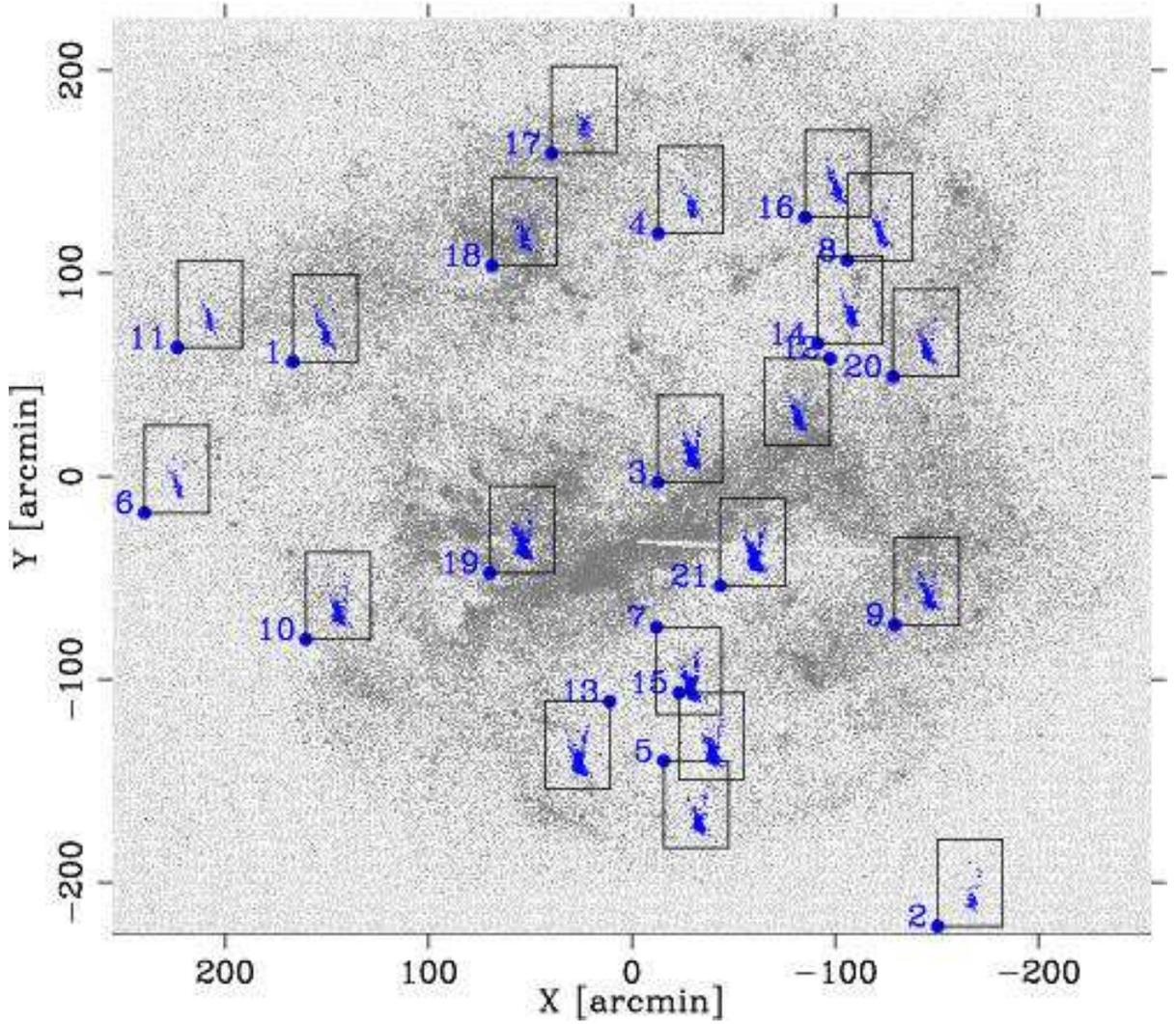}
\caption{Locations on the sky in a tangent plane projection and CMDs of
the 21 fields in the LMC superimposed on a map showing the distribution
of young stars from \citet{z04}.  North is up, east is to the left, and
the figure is centered at $(\alpha, \delta) = (5^h 18\fm 8, -68\degr
34\arcmin)$.  Each field location is marked with a filled circle.  All
of the CMDs have the same color and magnitude range, $-1 <
m_{606W}-m_{814W} < 2$ and $26 > m_{606W} > 14$, respectively.}
\label{fig:LMC_YS}
\end{figure}

Figure~\ref{fig:LMC_YS} shows the location of the LMC fields along with
the distribution of young stars as mapped by \citet{z04}.
The figure also shows the CMD for each field.  Most fields contain
both a young and an old stellar population.  Exceptions are
the fields in the northern spiral arm, which contain mostly a young
population, and field L2, which contains only an old population.

\clearpage

\begin{figure}[t!]
\centering
\includegraphics[angle=-90,scale=0.75]{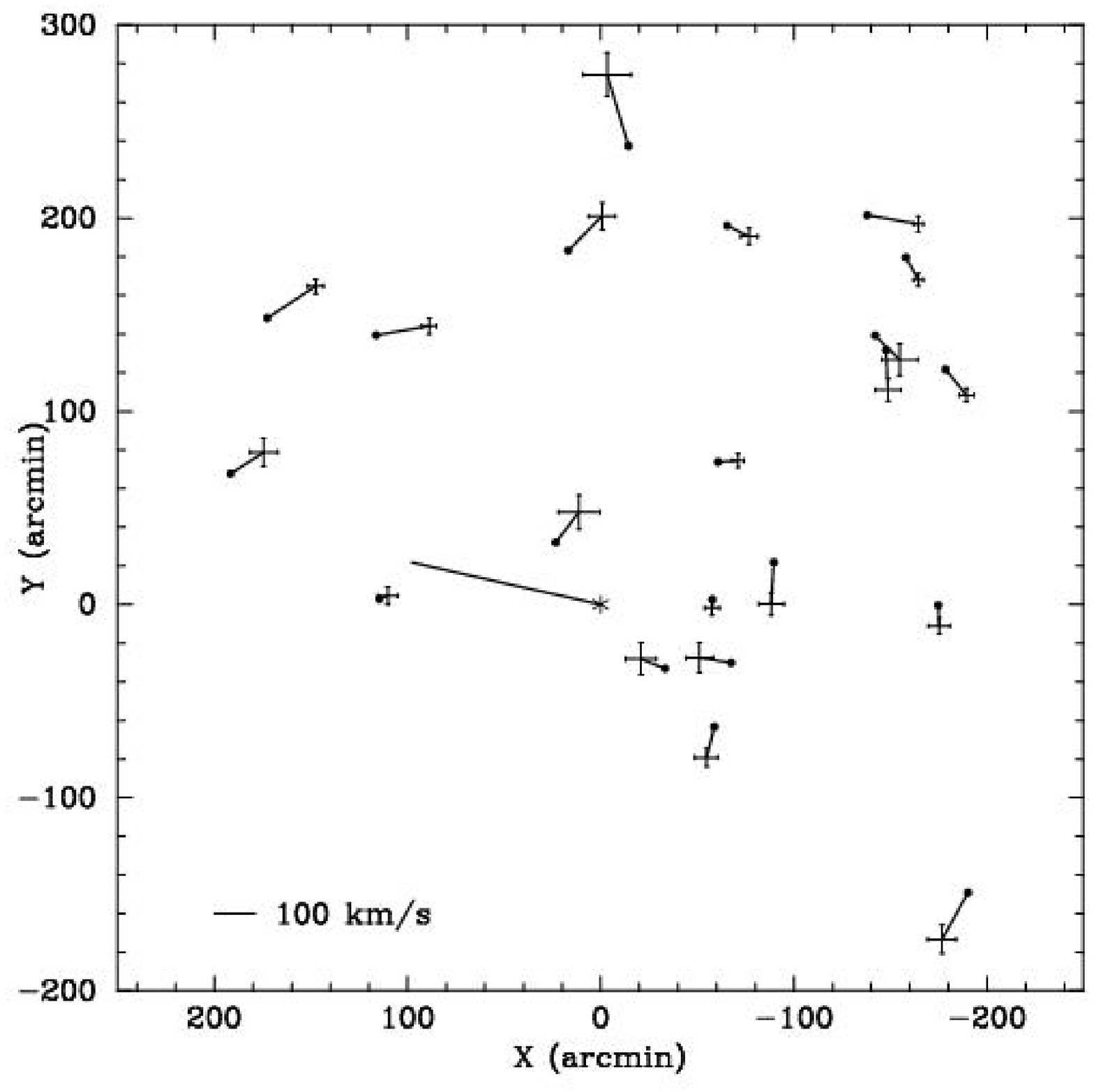}
\caption{Magnitude and direction of proper motions remaining after
subtracting the contributions due to the changing perspective of the
center-of-mass space velocity.  These proper motions contain
information about internal motions.  Each filled circle is at the
location on the sky of one of the 21 fields in the LMC in a tangent
plane projection.  North is up and east is to the left.  The line
emanating from each field location is the proper motion for that field
and the uncertainty is indicated by the error bars at its tip.  The
asterisk symbol marks the kinematical center of the LMC at $(\alpha,
\delta) = (5^h27\fm 6,-69\arcdeg 52.2\arcmin)$ and the line
originating from it has a length and direction proportional to the
adopted proper motion of the center of mass: $\mu_{\alpha,cm} =
195.6$~mas~century$^{-1}$ and $\mu_{\delta,cm} =
43.5$~mas~century$^{-1}$.  The line segment in the lower-left corner
shows a proper motion corresponding to a tangential velocity of
100~km~s$^{-1}$.  A visual inspection of the figure shows a clear
signature of a clockwise rotation.}
\label{fig:LMC_rot}
\end{figure} 

\clearpage

The LMC is known to exhibit rotation on the basis of the radial
velocities of HI and stars \citep[\textit{e.g.},][]{kim98,vdm02,om07}
and Figure~\ref{fig:LMC_YS} shows that the fields are distributed
widely in azimuth around the galaxy center.  Thus, the measured proper
motions listed in Table~1 must contain contributions from both the
center-of-mass space motion, including the effect of changing
perspective, and disk rotation.  They may also contain contributions
from the precession and nutation of the disk \citep{vdm02}, and from
streaming due to the bar or a tidal interaction.  To search for
internal motions we must remove the effect of the changing
perspective, which is calculable given a line-of-sight velocity and
proper motion of the galaxy center of mass and the galaxy distance
\citep[see][]{vdm02}.  Adopting values for these three quantities of
262.2~km~s$^{-1}$ \citep{vdm02}, our best estimate obtained as
described below (it depends slightly on the adopted rotation), and
50.1~kpc \citep{vdm02} yields the results in Figure~\ref{fig:LMC_rot}.
It plots at the location of each field the direction and magnitude of
the proper motion that arises only because of the internal motions of
the LMC.  Visual inspection shows a clear signature of a clockwise
rotation, albeit with superimposed noise.  The amplitude of the proper
motions for the fields farthest from the kinematical center implies
tangential velocities larger than 100~km~s$^{-1}$.  K06a noted a hint
of this rotational pattern in their equivalent Figure~12, but it was
not as clear as in Figure~\ref{fig:LMC_rot}.

\begin{figure}[t!]
\centering
\includegraphics[angle=-90,scale=0.75]{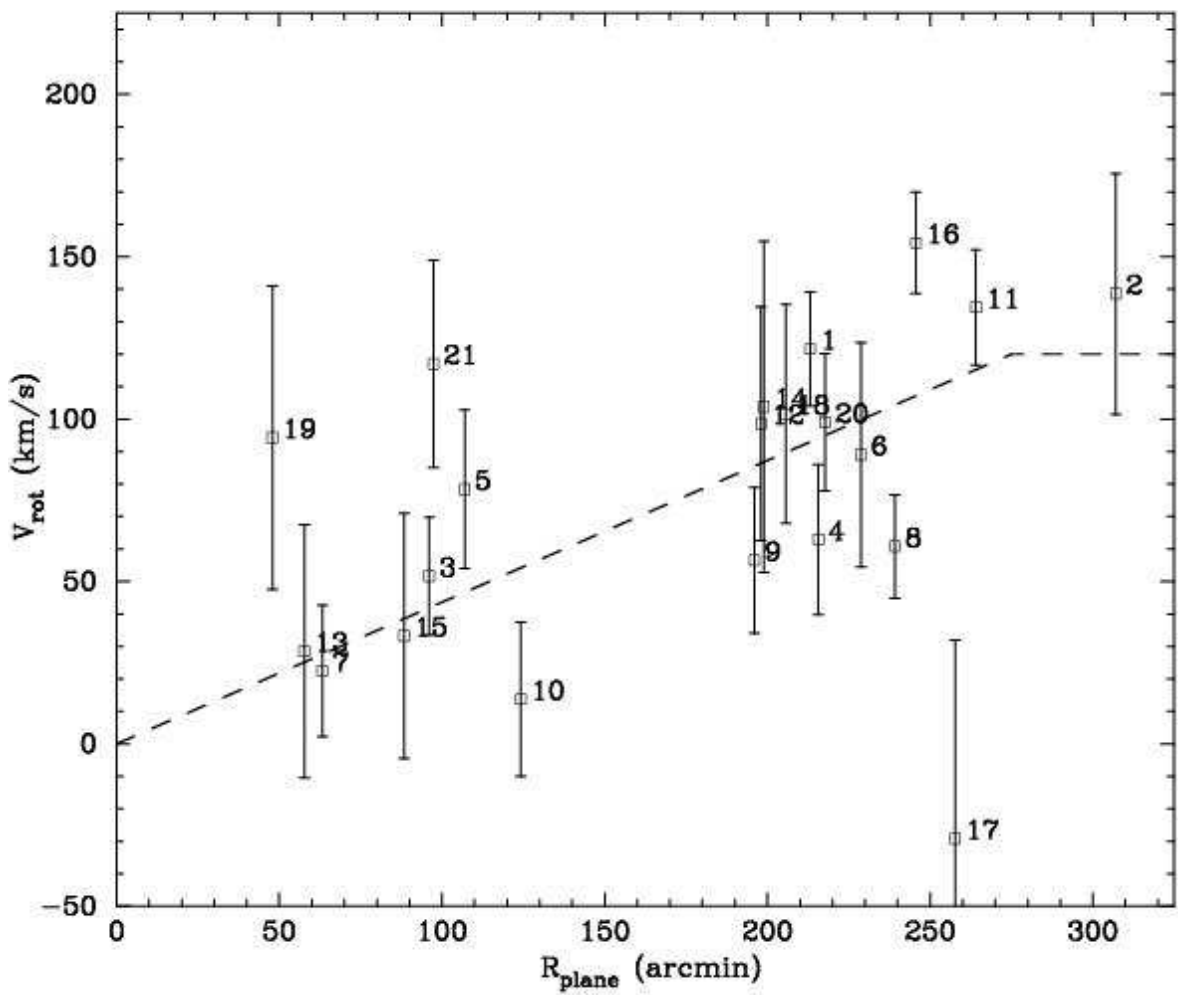}
\caption{Rotational velocity, $V_{rot}$, implied by a proper motion
that was corrected for changing perspective plotted as a function of
radius in the plane of the disk, $R_{plane}$.  This is a velocity in
the plane of the disk and perpendicular to the line of sight of a
stationary observer at the center of the LMC.  For easy reference,
each point is labeled with a field number.  The dashed curve is the
best-fitting model rotation curve assumed to be linearly increasing to
a radius of 275~arcmin and flat beyond; it has
$V_{275}=120$~km~s$^{-1}$.}
\label{fig:LMC_rc}
\end{figure}

Each proper motion in Figure~\ref{fig:LMC_rot} can be resolved into a
component along and perpendicular to the direction expected for
circular rotation in an inclined disk.  The first of these components
implies an amplitude for the rotation curve at a radius in the disk
plane.  The formulae for translating positions and proper motions in
the sky to radii and rotation velocities in the disk are given by
\citet{vdm02}.  Figure~\ref{fig:LMC_rc} plots the amplitude of the
rotation curve, $V_{rot}$, \textit{versus} radius in the plane of the
disk, $R_{plane}$, for each field.  The uncertainties are determined
from the uncertainties in the measured proper motions using
propagation of errors.  The calculations assume that the kinematical
center of the disk is at $(\alpha, \delta) = (5^h27\fm 6,-69\arcdeg
52.2\arcmin)$ and that the disk inclination and position angle of the
line of nodes are $34\fdg 7$ and $129\fdg 9$, respectively
\citep{vdm02}.  All of the fields except one have positive $V_{rot}$,
so the proper motions imply the presence of rotation.
Figure~\ref{fig:LMC_rc} shows that $V_{rot}$ increases with increasing
$R_{plane}$.  Some of the largest values of $V_{rot}$ are for L1, L11,
and L16, which are in the northern spiral arm.  This suggests that the
spiral arm has a motion different than that of the rest of the disk,
possibly because of a warp in the disk plane or because it is a tidal
tail.  However, other fields in the northern spiral arm, such as L4, L6,
and L18, have values of $V_{rot}$ similar to those of the rest of
the disk.

\clearpage

\clearpage

Estimates of the rotation of the LMC using radial velocities of carbon
stars \citep[K06a;][]{vdm02} and HI \citep{kim98,om07} find $V_{rot}$
increasing approximately linearly with $R_{plane}$ to a value of 60 --
80~km~s$^{-1}$ at a radius of about 275~arcmin (4.0~kpc) and roughly
constant beyond.  Figure~\ref{fig:LMC_rc} shows a larger amplitude for
the rotation.  We adopt a simple rotation curve that rises linearly to
a radius of 275~arcmin and is constant beyond.
Correcting the observed proper motions of each field for perspective
and rotation produces an estimate of the proper motion of the center
of mass.  The best estimates of the rotation curve and the center-of-mass
proper motion minimize the scatter of these estimates around their
weighted mean.  We add an additional uncertainty of
12.4~mas~century$^{-1}$ in quadrature to both components of the
measured proper motion of each field in order to produce a $\chi^2$
per degree of freedom of 1.0 for the best fit.  The result for the
center-of-mass proper motion and amplitude of the rotation curve at a
radius of 275~arcmin are
\begin{eqnarray}
\mu_{\alpha,cm} &=& 195.6\pm 3.6\ \textrm{mas century}^{-1} \label{eq:macm}\\
\mu_{\delta,cm} &=& 43.5 \pm 3.6\ \textrm{mas century}^{-1} \label{eq:mdcm}\\
V_{275} &=& 120 \pm 15\ \textrm{km s}^{-1}. \label{eq:vf}
\end{eqnarray}
These are our best estimates for these quantities.  The uncertainties
are derived by increasing $\chi^2$ by 1.0 above the minimum
\citep[\textit{e.g.},][]{p92} and so include the adopted additional
uncertainty.  Estimating the uncertainties in the right ascension and
declination components of the mean proper motion from the scatter of
the estimates around their weighted mean, as done by K06a, yields
4.1~mas~century$^{-1}$ and 4.5~mas~century$^{-1}$, respectively.  The
rotation curve implied by Equation~\ref{eq:vf} is shown as the dashed
curve in Figure~\ref{fig:LMC_rc}.

\clearpage

\begin{figure}[t!]
\centering
\includegraphics[angle=-90,scale=0.75]{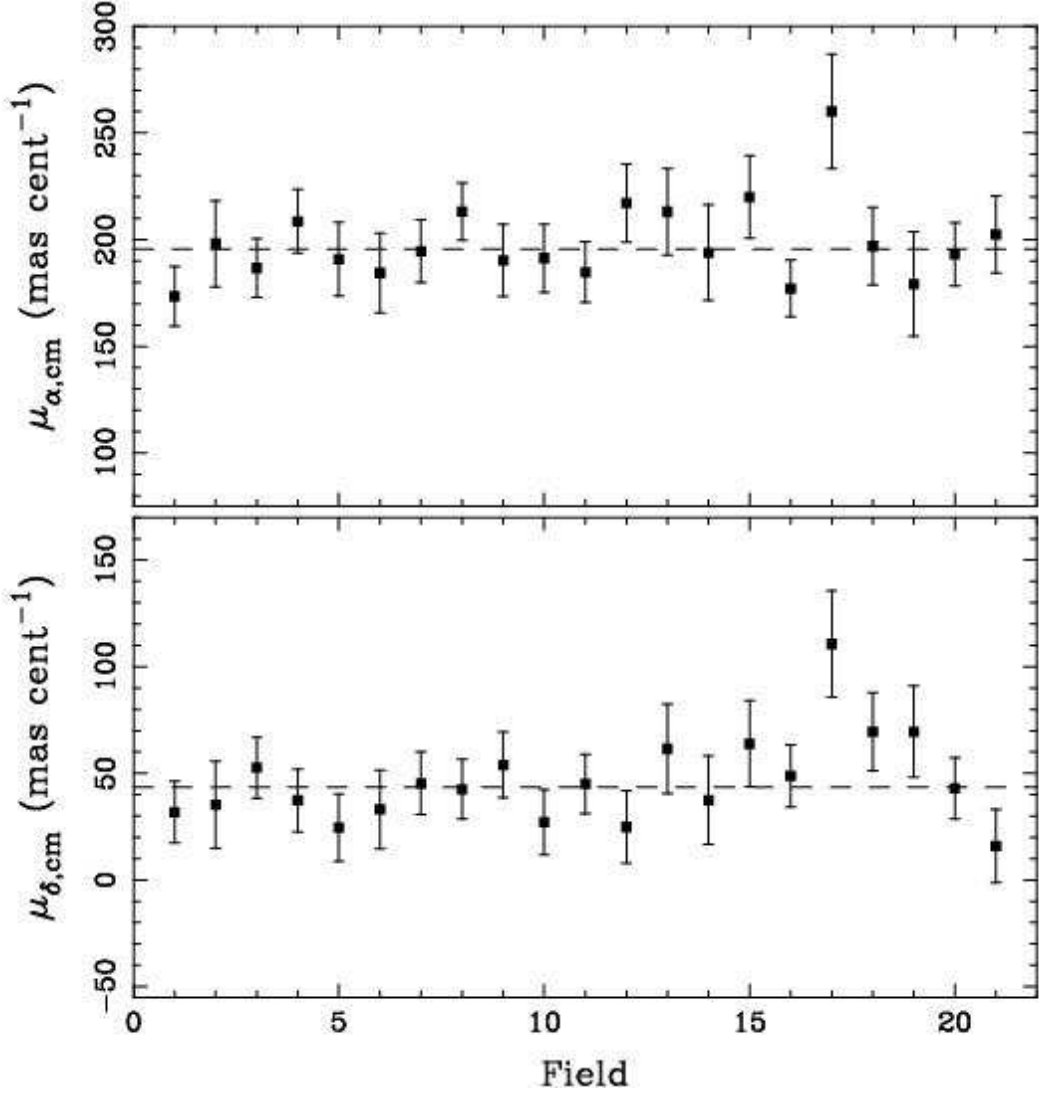}
\caption{Center-of-mass proper motion for the LMC determined by each of
21 fields as found in this article using $V_{275} = 120$~km~s$^{-1}$.
The error bars do not include the additional uncertainty discussed in
the text.  \textit{Top panel}: $\mu_{\alpha}$ \textit{versus} field
number.  \textit{Bottom panel}: $\mu_{\delta}$ \textit{versus} field
number.  The dashed horizontal lines are our weighted mean proper
motions for each component.  For easy comparison, both panels have the
same vertical scale, which is also the same as in
Figures~\ref{fig:LMCpmr} and \ref{fig:LMCpmc}.}
\label{fig:LMC_cpm}
\end{figure}

\clearpage

The difference between the rotation curves determined from the radial
velocities \citep[K06a;][]{kim98} and ours could be reduced by
decreasing the inclination of the disk.  A complete reanalysis would
simultaneously fit the radial velocity and proper motion data to
determine the rotation curve and orientation of the disk.  However,
such a fit is beyond the scope of this article.  Recently,
\citet{om07} used the K06a proper motion to study the internal motions
of the LMC implied by the radial velocities of the HI, carbon stars
(an intermediate age stellar population), and red supergiants (a young
stellar population).  They confirm the HI rotation curve and the other
spatially and kinematically distinct features first seen in the HI by
\citet{kim98} and \citet{ss03}.  The carbon stars share the kinematics
of the HI, but the rotation curve of the red supergiants rises to a
value of 107~km~s$^{-1}$, which is similar to what we find from the
proper motions and, in particular what we find for those fields in the
northern spiral
arm that are dominated by a young stellar population.  Some of the red
supergiants implying the largest rotation velocity are also in the
northern spiral arm: the magenta dots in Figure~2 of \citet{om07}.
Future proper motions with a longer time baseline may be able to
distinguish between the kinematics of different stellar populations.
Radial velocities for stars in the 21 fields would also help to compare
the rotation measured using radial velocities and proper motions.

Figure~\ref{fig:LMC_cpm} plots the center-of-mass proper motions for
each field derived with $V_{275} = 120$~km~s$^{-1}$ \textit{versus}
field number.  The dashed horizontal lines are the weighted means for
each component listed in Equations~\ref{eq:macm} and \ref{eq:mdcm}.
The scatter of the points around the weighted mean for each component
is smaller than the scatter in Figure~\ref{fig:LMCpmc}, which uses the
smaller rotation amplitude of K06a.  The significant reduction in the
scatter supports the larger amplitude for the rotation found in this
article.

\begin{figure}[t!]
\centering
\includegraphics[angle=-90,scale=0.75]{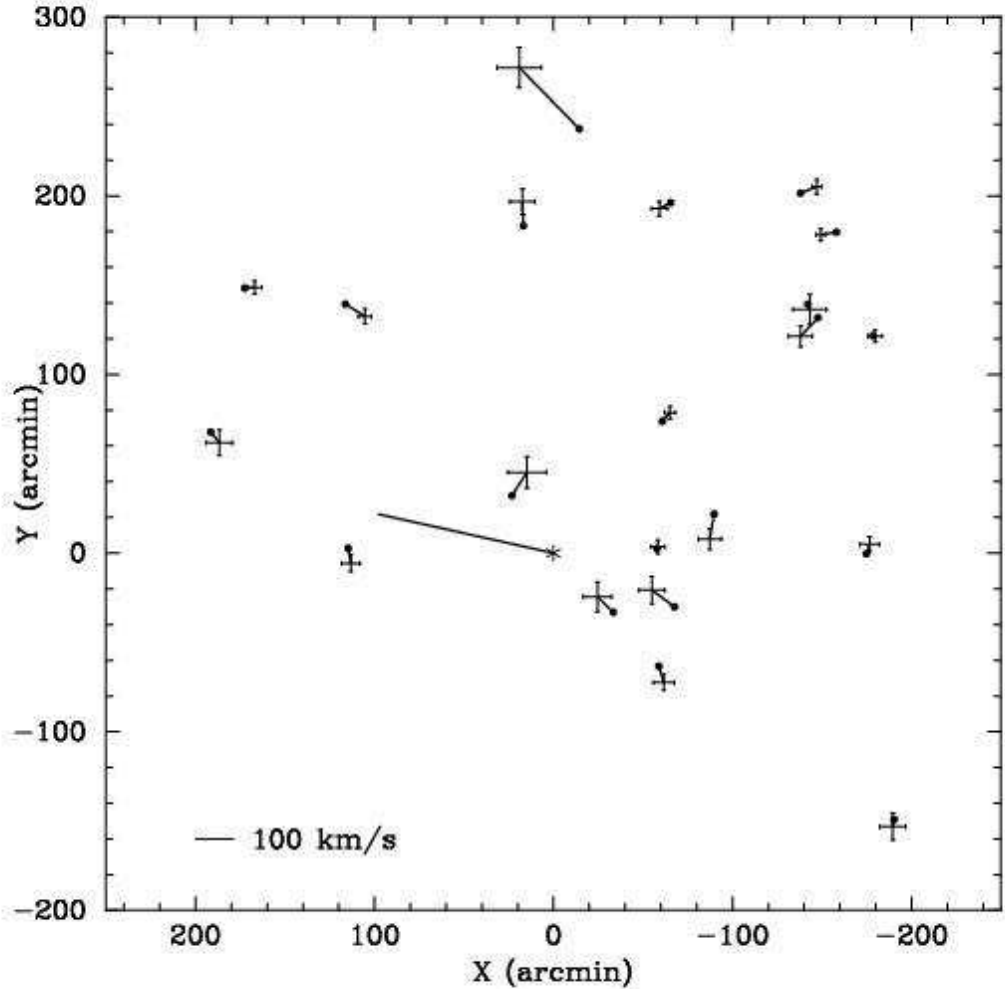}
\caption{Magnitude and direction of the proper motions remaining after
subtracting the contributions due to our best-fit rotation from the
proper motions in Figure~\ref{fig:LMC_rot}.  The resulting residual
vectors have directions and magnitudes determined by measurement
errors and by departures from the circular motions of the fitted
rotation curve.  The figure shows no clear pattern of streaming
motions.  The most significant residuals are found among the fields in
the northern spiral arm.}
\label{fig:LMC_res}
\end{figure}

The additional 12.4~mas~century$^{-1}$ of scatter in each component of
the measured proper motions found above implies the presence of some
combination of internal motions that depart from our adopted rotation
curve and errors larger than our measurement uncertainties.  Our
measurement uncertainties are derived from the scatter around the
best-fit coordinate transformation between epochs and should be
realistic in most cases.  An undetected systematic error might be
present if there is a gap between the high $S/N$ of the QSO and the
lower $S/N$ values of the stars.  Only fields L1, L3, and L11 have
such gaps.  L1 has a significant departure from the mean in
Figure~\ref{fig:LMC_cpm}, but L3 and L11 do not.  L17 has the largest
departure in Figure~\ref{fig:LMC_cpm} and shows strong trends with
$S/N$ and contains few stars, making correcting for those trends
difficult.  A proper motion of 12.4~mas~century$^{-1}$ corresponds to
a tangential velocity of 30~km~s$^{-1}$ and internal motions of this
size would indicate significant departures from circular motion.
Figure~\ref{fig:LMC_res} plots the proper motions remaining after
subtracting the contributions due to the rotation curve shown in
Figure~\ref{fig:LMC_rc} from the proper motions in
Figure~\ref{fig:LMC_rot}.  The figure does not show a clear pattern of
streaming motions.  The most significant residuals are found among the
fields in the northern spiral arm, but it is unclear what physical
mechanism could produce larger incoherent departures from the adopted
rotation curve there.  Again, it would be useful to obtain radial
velocities for stars in the proper motion fields.

\clearpage

\clearpage

Our proper motion for the center of mass of the LMC differs from that
of K06a by 7.4~mas~century$^{-1}$ in the right ascension component and
0.5~mas~century$^{-1}$ in the declination component.  The difference
in the right ascension components is as large as the uncertainty
quoted by K06a.  However, our proper motion confirms the surprising
result of K06a that led us to begin this investigation: the large
space velocity for the LMC.  The proper motion for the LMC found in
this article implies a galactocentric radial and tangential velocity
of $93.2\pm 3.7$~km~s$^{-1}$ and $346 \pm 8.5$~km~s$^{-1}$,
respectively.

\clearpage

\begin{figure}[h!]
\centering
\includegraphics[angle=0,scale=0.9]{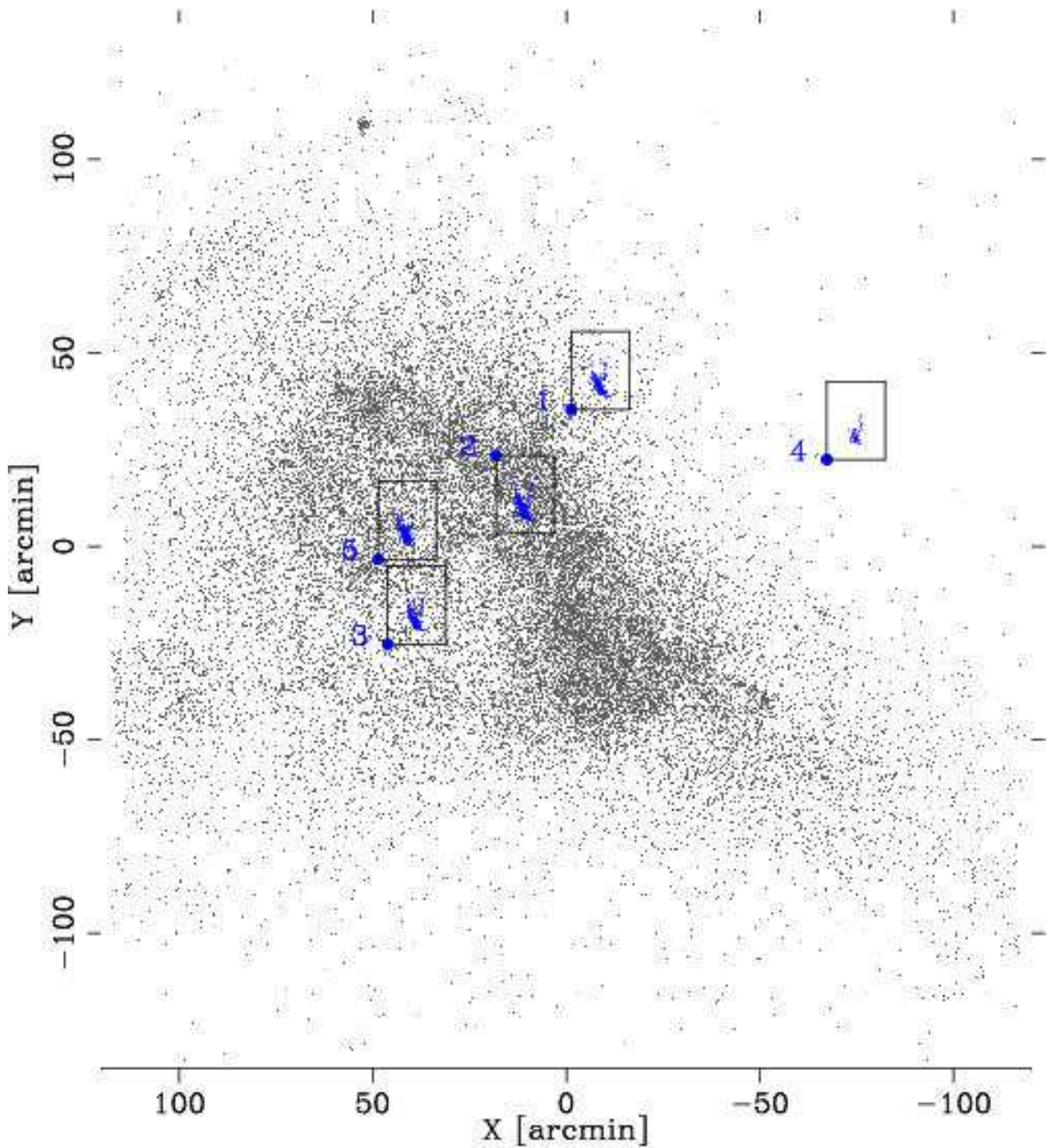}
\caption{Locations on the sky in a tangent plane projection and CMDs of
the five fields in the SMC superimposed on a map showing a
distribution of young stars from \citet{z00}.  North is up, east is to
the left and the figure is centered at $(\alpha, \delta) = (0^h 51\fm
6, -72\degr 52\arcmin)$.  Each field location is marked with a filled
circle.  All of the CMDs have the same color and magnitude range, $-1
< m_{606W}-m_{814W} < 2$ and $26 > m_{606W} > 14$, respectively.}
\label{fig:SMC_YS}
\end{figure}

\clearpage

\subsection{The SMC}
\label{sec:SMC}

Figure~\ref{fig:SMC_YS} shows the location of the SMC fields, their
CMDs, and the distribution of young stars as mapped by
\citet{z00}.  The figure shows that the surface density of young stars
at the location of S4 is lower than that for the other fields.  The
CMD for S4 contains mostly old stars, whereas the CMDs for the other
fields contain both old and young stars.

\begin{figure}[t!]
\centering
\includegraphics[angle=-90,scale=0.75]{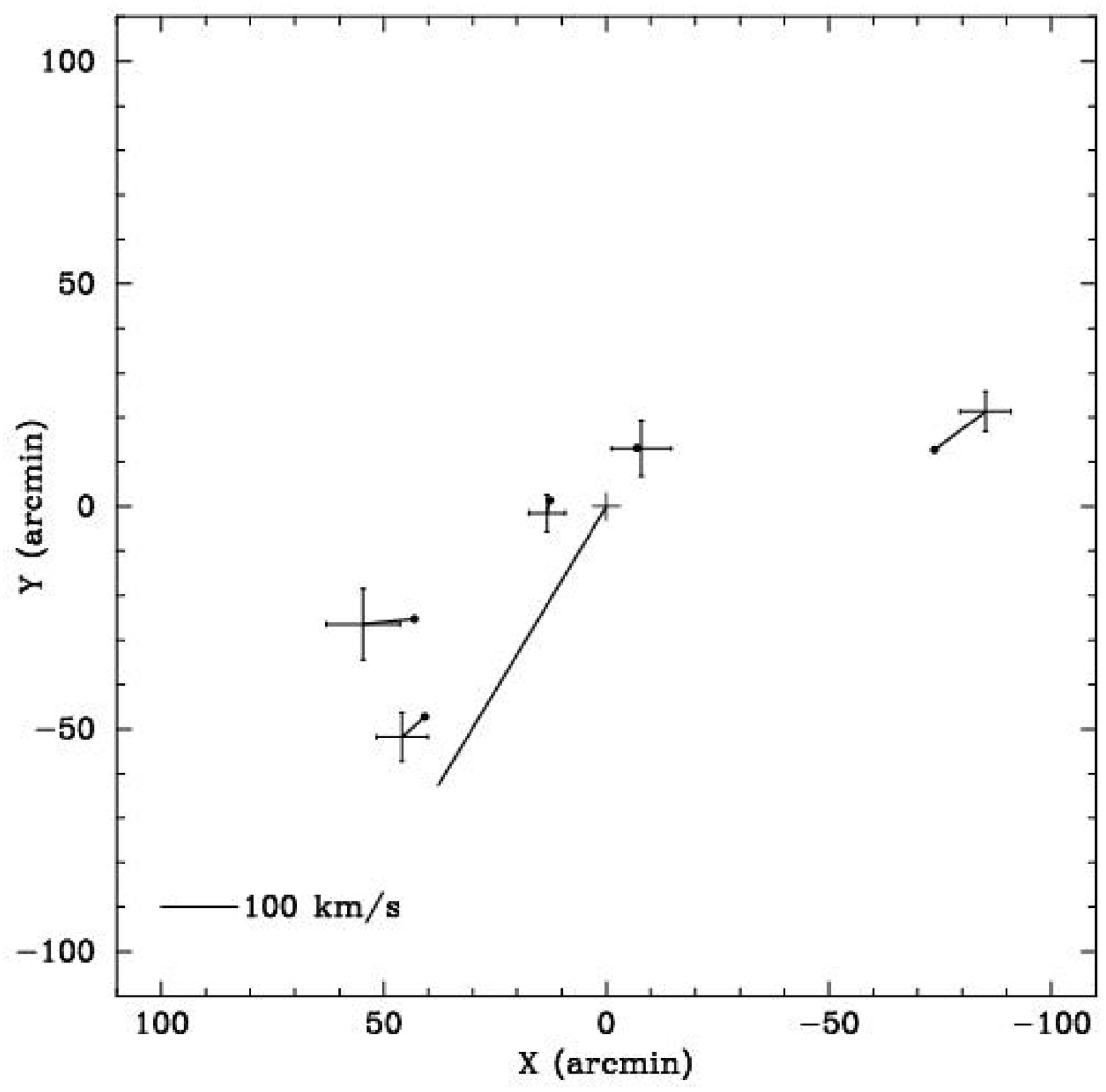}
\caption{Magnitude and direction of proper motions remaining after
subtracting the contributions due to the changing perspective of the
center-of-mass space velocity.  These proper motions contain
information about internal motions.  Each filled circle is at the
location on the sky of one of the five fields in the SMC in a tangent
plane projection.  North is up and east is to the left.   The line
emanating from each field location is the proper motion for that field
and the uncertainty is indicated by the error bars at its tip.  The
asterisk symbol marks the kinematical center of the SMC at
$(\alpha, \delta) = (0^h52\fm 8,-72\arcdeg 30\arcmin)$ and the line
originating from it has a length and direction proportional to the
adopted proper motion of the center of mass: $\mu_{\alpha,cm} =
80.8$~mas~century$^{-1}$ and $\mu_{\delta,cm} =
-125.6$~mas~century$^{-1}$.  The line segment in the lower-left corner
shows a proper motion corresponding to a tangential velocity of
100~km~s$^{-1}$.  A visual inspection of the figure shows no evidence
of rotation, but suggests the presence of radial streaming away from
the center.}
\label{fig:SMC_rot}
\end{figure}

Figure~\ref{fig:SMC_rot}, which is analogous to
Figure~\ref{fig:LMC_rot}, shows proper motions that were corrected for
the changing perspective of the center-of-mass space velocity of the
SMC.  We adopt a distance of 61.7~kpc \citep{c00}, our best estimate
of the proper motion of the galaxy, and a line-of-sight velocity of
146.0~km~s$^{-1}$ \citep{hz06}.  Visual inspection shows no clear
signature of rotation.  There is a suggestion of radial streaming
motions along a north-west --- south-east line.  However, measurements
of the proper motions in more fields are necessary to confirm the
presence of this streaming.

\clearpage

\clearpage

\begin{figure}[h]
\centering
\includegraphics[angle=-90,scale=0.75]{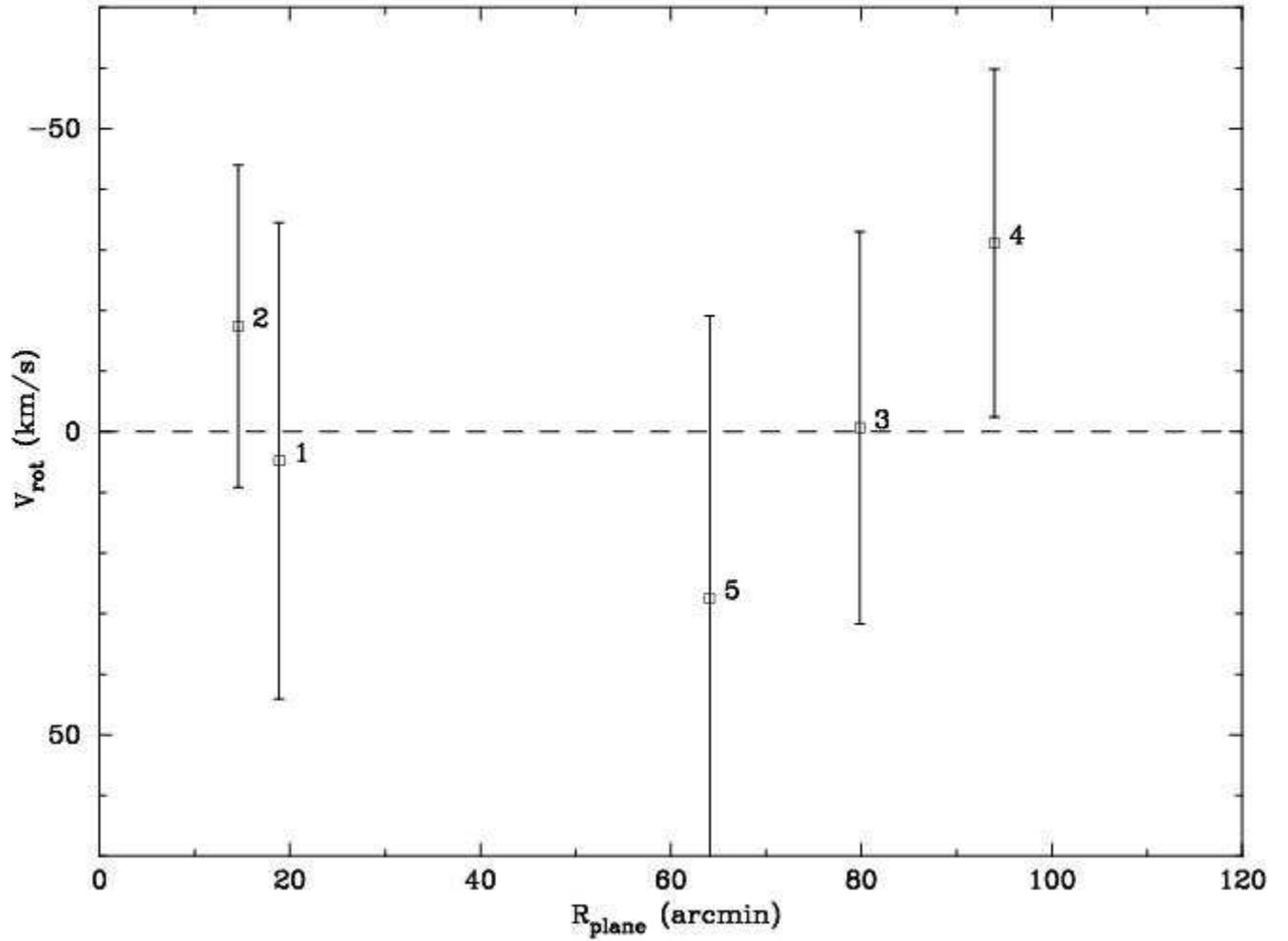}
\caption{Rotational velocity, $V_{rot}$, implied by a proper motion
that was corrected for changing perspective plotted as a function of
radius in the plane of the disk, $R_{plane}$.  This is a velocity in
the plane of the disk and perpendicular to the line of sight of a
stationary observer at the center of the SMC.  For easy reference, each
point is labeled with a field number.  There is no indication of rotation.}
\label{fig:SMC_rc}
\end{figure}

Figure~\ref{fig:SMC_rc}, which is analogous to Figure~\ref{fig:LMC_rc},
confirms that the circular velocities derived from the five fields are
consistent with no rotation.  The calculations assume that the
kinematical center of the disk is at $(\alpha, \delta) = (0^h52\fm
8,-72\arcdeg 30\arcmin)$ and that the disk inclination and position
angle of the line of nodes are $40\degr$ and $220\degr$, respectively
\citep{st04}.  The velocity gradient seen in the HI \citep{st04} and
red giants \citep{hz06}, if interpreted as rotation, would imply a
rotation curve in Figure~\ref{fig:SMC_rc} linearly rising to an
amplitude of $\pm$50~km~s$^{-1}$ at $R_{plane} = 120$~arcmin.  The
data in Figure~\ref{fig:SMC_rc} cannot rule out such a curve.

\clearpage

\clearpage

\begin{figure}[h]
\centering
\includegraphics[angle=-90,scale=0.75]{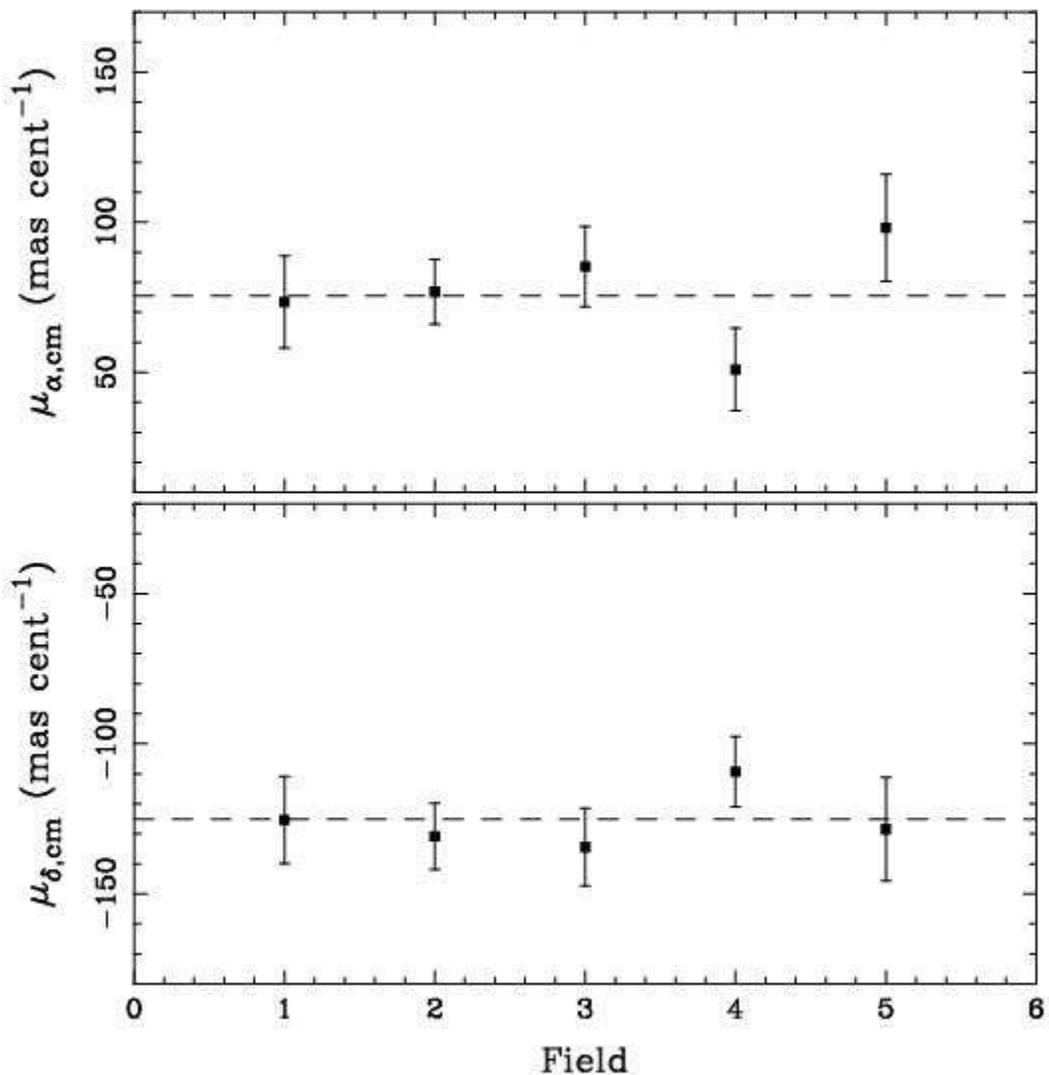}
\caption{Center-of-mass proper motion for the SMC determined by each of
five fields as found in this article assuming no rotation.  The error
bars do not include the additional uncertainty discussed in the text.
\textit{Top panel}: $\mu_{\alpha}$ \textit{versus} field number.
\textit{Bottom panel}: $\mu_{\delta}$ \textit{versus} field number.
The dashed horizontal lines are our weighted mean proper motions for
each component.  For easy comparison, both panels have the same
vertical scale, which is also the same as in Figure~\ref{fig:SMCpmr}.}
\label{fig:SMC_cpm}
\end{figure}

As for the LMC, each field yields a measurement of the center-of-mass
proper motion and these are plotted in Figure~\ref{fig:SMC_cpm}.  The
calculations assume no rotation.  The dashed lines are the weighted
means for each component and their values are
{\samepage
\begin{eqnarray}
\mu_{\alpha,cm} &=& 75.4\pm 6.1\ \textrm{mas century}^{-1}
\label{eq:smc-macm}\\
\mu_{\delta,cm} &=& -125.2 \pm 5.8\ \textrm{mas century}^{-1}.
\label{eq:smc-mdcm}
\end{eqnarray}}
These are our best estimates for these quantities.  We add an
additional uncertainty of 7.2~mas~century$^{-1}$ in quadrature to both
components of the measured proper motion of each field in order to
produce a $\chi^2$ per degree of freedom of 1.0 for the scatter around
the means.  Estimating the uncertainties in the right ascension and
declination components of the mean proper motion from the scatter of
the estimates around their weighted mean yields 7.0~mas~century$^{-1}$
and 3.9~mas~century$^{-1}$, respectively.  Our uncertainties for
$\mu_{\alpha,cm}$ and $\mu_{\delta,cm}$ are a factor of 3 smaller
than those of K06b.  The principal reason for our smaller uncertainties
is that we treat all five fields as independent measurements, whereas
K06b treated fields S1, S2, and S3 as a single measurement, S5 as
another, and excluded S4.  Our reanalysis, which corrects for the
effects caused by degrading CTE and for the trends of mean proper
motion with $S/N$, has reduced the systematic errors and, thus,
justifies treating these fields as independent.  The additional
uncertainty, added in quadrature to the measurement uncertainties
of the SMC, is similar to that for the LMC, which further supports our
quoted uncertainties.

\clearpage

\clearpage

Our proper motion for the center of mass of the SMC differs from that
of K06b by 40.6~mas~century$^{-1}$ in the right ascension component
and 8.2~mas~century$^{-1}$ in the declination component.  The
difference in the declination components is smaller than the
uncertainty quoted by K06b, whereas the difference in the right
ascension components is 2.3 times larger than the K06b uncertainty and
6.7 times larger than our uncertainty.  The difference in the right
ascension component arises because we include field S4 in the average,
have a lower value from field S5 because of the correction of trends
with $S/N$, and, as discussed above, treat all five measurements as
independent.

The proper motion for the SMC found in this article implies a
galactocentric radial and tangential velocity of $6.8\pm
2.4$~km~s$^{-1}$ and $259\pm 17$~km~s$^{-1}$, respectively.  The
relative velocity between the LMC and SMC is $142\pm 19$~km~s$^{-1}$,
which is 37~km~s$^{-1}$ higher than that found by K06b.

\section{Summary}
\label{sec:summary}

This article reports a reanalysis of images taken with the HRC of the
ACS on HST first analyzed by K06a and K06b to measure the proper
motions of the LMC and SMC.  Central to the method is the presence of a
QSO in a field; the proper motion derives from the reflex motion of a
QSO with respect to the stars of the galaxy.  There are 21 fields in
the LMC and five in the SMC.  The key findings and conclusions from our
analysis are:

\begin{enumerate}
\item We have detected a trend between the mean measured motion
and the brightness of objects that is present to a varying degree in a
majority of the fields.  We are unable to identify the source of these
trends.  If not accounted for, the trend can significantly affect the
measured proper motion.  Because the QSO is one of the brightest
objects in the field, we minimize the influence of the trends by
restricting the sample of stars contributing to the measurement of the
proper motion in a field to those whose $S/N$ is above some limit.
Proper motions derived with a wider range for the $S/N$ of the sample
agree better with those of K06a and K06b, thus arguing that the trends are
present in their analyses too.

\item  Our analysis also approximately corrects the effects caused by the
decreasing charge transfer efficiency with time in the CCD of the HRC.
These corrections are smaller than those for the trends with $S/N$.

\item  For most of the fields in the LMC and SMC, our measured proper
motion agrees within the quoted uncertainties with that of K06a or
K06b.  In those cases where the measurements differ (notably fields L13,
L15, L21, and S5), the difference is due to our measurements being
corrected for the trends with $S/N$.  Our measured proper motions for
the 21 fields in the LMC and the five fields in the SMC show less
scatter around the two mean center-of-mass proper motions.  With our
improved analysis, it is no longer necessary to exclude any of the
fields from the calculations of the means.

\item  Removing a contribution to the measured proper motions from the
changing perspective of the space velocity gives proper motion vectors
that contain information about the internal motions of the LMC.
Plotting these vectors on the sky shows a pattern of clockwise
rotation.  Converting each vector into an estimate of the rotation
velocity at a radius in the disk plane shows that the rotation of the
LMC has been clearly detected from the proper motions for the first
time.  Assuming a model rotation curve that rises linearly to a radius
of 275~arcmin and that is flat beyond yields a best-fit amplitude at
this radius of $120\pm 15$~km~s$^{-1}$.

\item  Our best estimate of the mean center-of-mass proper motion of
the LMC is $(\mu_{\alpha}, \mu_{\delta}) =
(195.6\pm 3.6, 43.5 \pm 3.6)$~mas century$^{-1}$.

\item  We do not detect rotation in the SMC.  The proper motions
suggest the presence of radial expansion, however more fields and
more precise measurements are needed to confirm the reality of
these streaming motions.

\item  Our best estimate of the mean center-of-mass proper motion of the SMC
is $(\mu_{\alpha}, \mu_{\delta}) = (75.4\pm 6.1, -125.2 \pm 5.8)$~mas
century$^{-1}$.  The uncertainties are 2.5 times smaller than those of
K06b because the improved internal consistency of our proper motions
permits treating all five fields as independent measurements.
\end{enumerate}

\acknowledgments

We thank the referee, Steven Majewski, and also Jay Anderson, Nitya
Kallivayalil, Roeland van der Marel, and Gurtina Besla for helpful
suggestions and discussions which made this article better.  We further
thank Jay Anderson for sending us his program to correct for the
geometric distortion in the HRC.  CP and SP acknowledge the financial
support of the Space Telescope Science Institute through the grants
HST-AR-10971 and HST-GO-10229 and of the National Science Foundation
through grant AST-0098650.  EWO acknowledges support from the Space
Telescope Science Institute through the grants HST-GO-07341.01-A and
HST-GO-08286.01-A and from the National Science Foundation through the
grants AST-0205790 and AST-0507511.

\clearpage



\begin{deluxetable}{lcrrrrrrr}
\tablecolumns{9} 
\tabletypesize{\small}
\tablewidth{0pc} 
\tablecaption{Comparison of Measured Proper Motions for the LMC}
\tablehead{&&&\multicolumn{2}{c}{This Article} &
\multicolumn{2}{c}{Kallivayalil et al.}&& \\
\colhead{Field}&$S/N$&N&\colhead{$\mu_{\alpha}$}&\colhead{$\mu_{\delta}$}&
\colhead{$\mu_{\alpha}$}&\colhead{$\mu_{\delta}$}&
\colhead{$\Delta\mu_{\alpha}$}&\colhead{$\Delta\mu_{\delta}$}\\
\colhead{(1)}&
\colhead{(2)}&
\colhead{(3)}&
\colhead{(4)}&
\colhead{(5)}&
\colhead{(6)}&
\colhead{(7)}&
\colhead{(8)}&
\colhead{(9)}}
\startdata
L1 &25& 25&$139.6\pm 7.4$&$ 65.9\pm 8.3$&$116.2\pm12.1$&$ 80.0\pm 8.9$&$ 23.4\pm14.2$&$-14.1\pm12.2$\\
L2 &25& 19&$223.3\pm14.6$&$-35.8\pm14.8$&$222.0\pm 7.0$&$-27.4\pm 6.7$&$  1.3\pm16.2$&$ -8.4\pm16.3$\\
L3 &25& 86&$179.0\pm 6.1$&$ 33.8\pm 7.0$&$197.6\pm 8.2$&$ 41.3\pm 6.1$&$-18.6\pm10.2$&$ -7.5\pm 9.3$\\
L4 &25& 22&$180.5\pm 8.8$&$ 17.8\pm 8.4$&$212.1\pm15.0$&$ 52.1\pm11.8$&$-31.6\pm17.4$&$-34.3\pm14.5$\\
L5 &25& 43&$204.1\pm11.6$&$  3.2\pm 9.0$&$205.4\pm 8.6$&$ 10.2\pm 9.7$&$ -1.3\pm14.4$&$ -7.0\pm13.3$\\
L6 &25& 14&$153.2\pm14.7$&$ 91.9\pm14.4$&$165.8\pm19.1$&$ 98.3\pm29.8$&$-12.6\pm24.1$&$ -6.4\pm33.1$\\
L7 &30& 93&$198.0\pm 7.7$&$ 25.5\pm 7.7$&$206.8\pm 7.3$&$ 42.6\pm 6.4$&$ -8.8\pm10.6$&$-17.1\pm10.0$\\
L8 &50& 16&$191.4\pm 5.2$&$ -7.0\pm 6.7$&$196.1\pm 8.1$&$ -5.0\pm 5.8$&$ -4.7\pm 9.6$&$ -2.0\pm 8.8$\\
L9 &25& 58&$199.2\pm10.9$&$ -6.3\pm 8.5$&$202.3\pm 7.7$&$ -4.7\pm 8.4$&$ -3.1\pm13.4$&$ -1.6\pm11.9$\\
L10&25& 60&$180.9\pm10.4$&$ 64.8\pm 9.2$&$193.4\pm 9.9$&$ 60.0\pm 8.7$&$-12.5\pm14.4$&$  4.8\pm12.7$\\
L11&25& 15&$141.6\pm 8.0$&$ 96.4\pm 7.4$&$108.9\pm35.2$&$118.6\pm18.7$&$ 32.7\pm36.1$&$-22.2\pm20.1$\\
L12&50& 14&$200.9\pm13.4$&$-23.2\pm11.6$&$244.7\pm16.9$&$ -2.4\pm18.3$&$-43.8\pm21.5$&$-20.8\pm21.7$\\
L13&50& 25&$221.1\pm15.9$&$ 48.7\pm16.6$&$245.5\pm13.4$&$116.8\pm 7.8$&$-24.4\pm20.8$&$-68.1\pm18.4$\\
L14&25& 37&$177.8\pm18.7$&$ -6.2\pm16.5$&$183.7\pm17.4$&$ 13.6\pm22.5$&$ -5.9\pm25.5$&$-19.8\pm27.9$\\
L15&50& 19&$229.9\pm14.4$&$ 38.1\pm15.6$&$273.8\pm17.3$&$ 57.3\pm21.7$&$-43.9\pm22.5$&$-19.2\pm26.7$\\
L16&25& 33&$151.8\pm 5.3$&$  9.7\pm 8.0$&$152.5\pm14.1$&$ 27.7\pm11.7$&$ -0.7\pm15.1$&$-18.0\pm14.2$\\
L17&25& 16&$224.8\pm24.8$&$108.9\pm22.5$&$231.3\pm28.9$&$100.4\pm18.5$&$ -6.5\pm38.1$&$  8.5\pm29.1$\\
L18&25& 24&$165.1\pm14.0$&$ 76.3\pm14.2$&$177.9\pm13.3$&$ 91.4\pm12.8$&$-12.8\pm19.3$&$-15.1\pm19.1$\\
L19&25&104&$171.4\pm21.3$&$ 77.6\pm17.7$&$173.5\pm16.3$&$ 84.4\pm20.8$&$ -2.1\pm26.8$&$ -6.8\pm27.3$\\
L20&25& 51&$181.3\pm 7.9$&$-12.9\pm 7.0$&$181.3\pm 7.1$&$  0.5\pm11.6$&$  0.0\pm10.7$&$-13.4\pm13.6$\\
L21&25&115&$202.2\pm13.0$&$-14.6\pm11.7$&$246.4\pm11.5$&$  5.4\pm12.2$&$-44.2\pm17.3$&$-20.0\pm16.9$\\
&&&&&& Average: & $-10.48$&$-14.69$\\
&&&&&& \textit{rms}: &  $19.65$& $15.73$\\
\enddata
\tablecomments{Proper motions are all in milli-arcseconds century$^{-1}$.}
\end{deluxetable}

\begin{deluxetable}{lrrrr}
\tablecolumns{5} 
\tablewidth{0pc} 
\tablecaption{Comparison of Center-of-Mass Proper Motions for the LMC}
\tablehead{&\multicolumn{2}{c}{This Article} &
\multicolumn{2}{c}{Kallivayalil et al.} \\
\colhead{Field}&\colhead{$\mu_{\alpha}$}&\colhead{$\mu_{\delta}$}&
\colhead{$\mu_{\alpha}$}&\colhead{$\mu_{\delta}$} \\
\colhead{(1)}&
\colhead{(2)}&
\colhead{(3)}&
\colhead{(4)}&
\colhead{(5)}}
\startdata
L1 &$156.8\pm 7.4$&$ 40.9\pm 8.3$&$133.4\pm12.1$&$ 55.0\pm 8.9$\\
L2 &$211.1\pm14.6$&$ 15.4\pm14.8$&$209.8\pm 7.0$&$ 23.8\pm 6.7$\\
L3 &$178.3\pm 6.1$&$ 47.8\pm 7.0$&$196.9\pm 8.2$&$ 55.3\pm 6.1$\\
L4 &$190.2\pm 8.8$&$ 34.9\pm 8.4$&$221.8\pm15.0$&$ 69.2\pm11.8$\\
L5 &$199.4\pm11.6$&$ 16.4\pm 9.0$&$200.7\pm 8.6$&$ 23.4\pm 9.7$\\
L6 &$173.2\pm14.7$&$ 47.9\pm14.4$&$185.8\pm19.1$&$ 54.3\pm29.8$\\
L7 &$195.7\pm 7.7$&$ 37.2\pm 7.7$&$204.5\pm 7.3$&$ 54.3\pm 6.4$\\
L8 &$197.1\pm 5.2$&$ 30.9\pm 6.7$&$201.8\pm 8.1$&$ 32.9\pm 5.8$\\
L9 &$193.6\pm10.9$&$ 39.0\pm 8.5$&$196.7\pm 7.7$&$ 40.6\pm 8.4$\\
L10&$188.9\pm10.4$&$ 38.5\pm 9.2$&$201.4\pm 9.9$&$ 33.7\pm 8.7$\\
L11&$163.2\pm 8.0$&$ 60.5\pm 7.4$&$130.5\pm35.2$&$ 82.7\pm18.7$\\
L12&$204.0\pm13.4$&$ 13.0\pm11.6$&$247.8\pm16.9$&$ 33.8\pm18.3$\\
L13&$219.7\pm15.9$&$ 54.9\pm16.6$&$244.1\pm13.4$&$123.0\pm 7.8$\\
L14&$181.5\pm18.7$&$ 28.6\pm16.5$&$187.4\pm17.4$&$ 48.4\pm22.5$\\
L15&$226.5\pm14.4$&$ 52.7\pm15.6$&$270.4\pm17.3$&$ 71.9\pm21.7$\\
L16&$159.0\pm 5.3$&$ 42.9\pm 8.0$&$159.7\pm14.1$&$ 60.9\pm11.7$\\
L17&$238.4\pm24.8$&$115.2\pm22.5$&$244.9\pm28.9$&$106.7\pm18.5$\\
L18&$178.0\pm14.0$&$ 74.6\pm14.2$&$190.8\pm13.3$&$ 89.7\pm12.8$\\
L19&$172.3\pm21.3$&$ 74.0\pm17.7$&$174.4\pm16.3$&$ 80.8\pm20.8$\\
L20&$183.3\pm 7.9$&$ 30.7\pm 7.0$&$183.3\pm 7.1$&$ 44.1\pm11.6$\\
L21&$199.2\pm13.0$&$  5.4\pm11.7$&$243.4\pm11.5$&$ 25.4\pm12.2$\\
\enddata
\tablecomments{Proper motions are all in milli-arcseconds century$^{-1}$.}
\end{deluxetable}

\begin{deluxetable}{lccrrrrrr}
\tabletypesize{\small}
\tablecolumns{9} 
\tablewidth{0pc} 
\tablecaption{Comparison of Measured Proper Motions for the SMC}
\tablehead{&&&\multicolumn{2}{c}{This Article} &
\multicolumn{2}{c}{Kallivayalil et al.}&& \\
\colhead{Field}&$S/N$&N&\colhead{$\mu_{\alpha}$}&\colhead{$\mu_{\delta}$}&
\colhead{$\mu_{\alpha}$}&\colhead{$\mu_{\delta}$}&
\colhead{$\Delta\mu_{\alpha}$}&\colhead{$\Delta\mu_{\delta}$}\\
\colhead{(1)}&
\colhead{(2)}&
\colhead{(3)}&
\colhead{(4)}&
\colhead{(5)}&
\colhead{(6)}&
\colhead{(7)}&
\colhead{(8)}&
\colhead{(9)}}
\startdata
S1 &25& 36&$  72.5\pm 13.5$&$-125.6\pm 12.5$&$  86.0\pm 11.3$&$-113.6\pm  9.5$&$ -13.5\pm 17.6$&$ -12.0\pm 15.7$\\
S2 &25& 90&$  78.3\pm  8.1$&$-130.3\pm  8.3$&$  82.5\pm  7.3$&$-120.8\pm  7.6$&$  -4.2\pm 10.9$&$  -9.5\pm 11.3$\\
S3 &25& 61&$  91.1\pm 11.5$&$-132.2\pm 10.9$&$ 102.2\pm  9.1$&$-120.1\pm 10.9$&$ -11.1\pm 14.7$&$ -12.1\pm 15.4$\\
S4 &25& 11&$  43.8\pm 11.5$&$-111.0\pm  9.0$&$  30.3\pm  7.3$&$ -86.6\pm 17.7$&$  13.5\pm 13.6$&$ -24.4\pm 19.9$\\
S5 &50& 17&$ 103.9\pm 16.5$&$-125.5\pm 15.9$&$ 147.1\pm 10.8$&$-114.3\pm 13.0$&$ -43.2\pm 19.7$&$ -11.2\pm 20.5$\\
&&&&&& Average: & $-11.69$&$-13.84$\\
&&&&&& \textit{rms:} & 20.57&  5.99\\
\enddata
\tablecomments{Proper motions are all in milli-arcseconds century$^{-1}$.}
\end{deluxetable}

\end{document}